\documentclass[10pt,aps,twocolumn,prd,superscriptaddress,showpacs,nofootinbib,noshowkeys,longbibliography,preprintnumbers]{revtex4-1}
\usepackage[dvips]{graphics,graphicx}
\usepackage{color}
\usepackage{amsmath, amssymb}
\usepackage{multirow}
\usepackage{longtable}
\usepackage[normalem]{ulem}  

\begin{document}
\title{Stable Yang-Lee zeros in truncated fugacity series from net-baryon number multiplicity distribution}
\date{\today}
\author{Kenji Morita}
\email{kmorita@yukawa.kyoto-u.ac.jp}
\affiliation{Yukawa Institute for Theoretical Physics, Kyoto University,
Kyoto 606-8502, Japan}
\affiliation{Frankfurt Institute for Advanced
  Studies,Ruth-Moufang-Strasse 1, D-60438, Frankfurt am Main, Germany}
\affiliation{Institute of Theoretical Physics, University of Wroclaw,
PL-50204 Wroc\l aw, Poland}
\author{Atsushi Nakamura}
\affiliation{Research Center for Nuclear Physics (RCNP), Osaka University, Ibaraki,
Osaka, 567-0047, Japan}
\affiliation{Theoretical Research Division, Nishina Center, RIKEN, Wako, Saitama
351-0198, Japan}
\preprint{YITP-15-44, RIKEN-QHP-190}
\begin{abstract}
 We investigate Yang-Lee zeros of grand partition functions as truncated
 fugacity polynomials of which coefficients are given by  the canonical
 partition functions $Z(T,V,N)$ up to $N \leq N_{\text{max}}$. 
 Such a partition function can be inevitably obtained from the
 net-baryon number  multiplicity distribution in relativistic heavy ion
 collisions, where the number of the event beyond $N_{\text{max}}$ has
 insufficient  statistics, as well as canonical approaches in lattice
 QCD. We use a chiral random matrix model as a solvable model for chiral
 phase transition in QCD and show that the closest edge of the
 distribution to real chemical potential axis is stable against cutting
 the tail of the multiplicity distribution. The similar behavior is also
 found in lattice QCD at finite temperature for Roberge-Weiss
 transition. In contrast, such a stability is found to be absent in the
 Skellam distribution which does not have phase transition.
 We compare the number of $N_{\text{max}}$ to obtain the stable Yang-Lee
 zeros with those of critical higher order cumulants.
\end{abstract}
\pacs{12.38.Gc, 12.38.Mh, 25.75.Nq, 25.75.Gz}
\maketitle


\section{Introduction}
Phase transition in quantum chromodynamics (QCD) is one of the central subjects in high energy
nuclear physics both theoretically and experimentally. First
principle lattice QCD (LQCD) calculations have shown that the transition from
quark-gluon plasma (QGP) to hadronic matter is of crossover type
at physical quark masses \cite{aoki06}, in which order parameters and
thermodynamic quantities change smoothly as functions of temperature.
At finite baryon density, one expects that the nature of the transition
can change. Unfortunately, little is known about the state of matter at
high baryon density from LQCD calculations
because of the difficulty in numerical simulation at finite baryon
chemical potential $\mu$
\cite{muroya03:_lattic_qcd,forcrand09:_simul_qcd_at_finit_densit}. 
Various approximation methods applied so far seem to work only $\mu < T$ or
a small volume or heavy quark mass region. Nevertheless, effective
models which implement relevant symmetries in QCD and large $N_c$
studies have shown that rich phase structure exists in high density
\cite{fukushima11:_phase_diagr_of_dense_qcd,fukushima13:_phase_diagr_of_nuclear_and}. In particular, if there is
a first order phase transition at $T=0$ and large $\mu$, there must be
a critical point (CP) at which the first order phase transition line
terminates and the transition becomes second order. 
Existence of CP is supported by many chiral effective models
\cite{asakawa89:_chiral_restor_at_finit_densit_and_temper},
but its location depends on the detail of the models \cite{stephanov04:_qcd}.

Stimulated by these theoretical results, the first beam energy scan
program at Relativistic Heavy Ion Collider (RHIC) has been carried out
in search for the CP \cite{xu14:_star,soltz14:_phenix}.
 Since  lower colliding energies leaves the incident
nucleons in the central region, one expects to explore higher baryon
density region at lower energies. There are a number of observables
which might have potential to indicate the transition from QGP to
hadronic matter. Among them, event-by-event fluctuations of conserved charges are
intimately connected to critical behavior associated with the phase
transition
\cite{stephanov98:_signat_of_tricr_point_in_qcd,stephanov99:_event_by_event_fluct_in,asakawa00:_fluct_probes_of_quark_decon,jeon00:_charg_partic_ratio_fluct_as}. Measurements
of the net-proton number fluctuations as a
proxy of the net-baryon one \cite{hatta03:_proton_number_fluct_as_signal}
and net electric charges have
been presented for Au+Au collisions at various energies from
$\sqrt{s_{NN}}=7.7$ GeV to 200 GeV
\cite{aggarwal10:_higher_momen_of_net_proton,STAR_pn_2013,adamczyk14:_beam_energ_depen_of_momen}. Given
the fact that multiplicity of different particle species are well
described by statistical models
\cite{Statmodelreview_QGP3,andronic06:_hadron,cleymans06:_compar}, one may regard event-by-event fluctuations 
of conserved charges as those of the grand canonical ensembles at
chemical freeze-out temperature $T$ and baryonic chemical potential
$\mu$. Through
systematic analyses of the location of
$(T,\mu)$ corresponding to each colliding systems \cite{cleymans06:_compar}, 
one can map experimental measurements for property of the matter on
$T-\mu$ plane. 
Furthermore, recent LQCD results at physical quark masses indicate
that the crossover region coincides with the chemical freeze-out at least
for $\mu < T$ \cite{allton05:_therm_of_two_flavor_qcd,kaczmarek11:_phase_qcd}.
One may look for  remnant of the chiral criticality in the crossover region,
originating from second order phase transition in the vanishing quark
mass \cite{karsch11:_probin_freez_out_condit_in,friman11:_fluct_as_probe_of_qcd,morita14:_critic_net_baryon_number_probab}.

Property of the transition can be characterized by behavior of
fluctuations of conserved charges as well as an order
parameter and its fluctuations
\cite{hatta03:_univer_qcd_critic_and_tricr,koch:_hadron_fluct_and_correl}. In
the case of the chiral phase
transition in QCD, the chiral order parameter or quark-antiquark
condensate $\langle\bar{q}q\rangle$ couples to quarks carrying the baryon
number and the electric charge. Thus, the second order chiral phase
transition in the chiral limit at finite temperature is characterized by
not only divergent fluctuation of the order parameter but also higher
order cumulants of the net baryon number and the net electric charge \cite{stephanov09:_non_gauss_fluct_near_qcd_critic_point}.
The divergence of the conserved charge fluctuations is governed by the
critical exponent of the specific heat $\alpha$ which depends on the
universality class QCD belongs to. Although it is not completely
determined yet \cite{aoki12:_chiral_symmet_restor_eigen_densit,bhattacharya14:_qcd},
recent simulations \cite{ejiri09:_magnet_equat_of_state_in_flavor_qcd}
indicate $O(4)$ in the three dimensions, as conjectured by Pisarski and
Wilczek \cite{pisarski84:_remar_chiral_trans_in_chrom}.
In this case, the first divergent cumulant appears at the sixth order.
At finite but small quark masses, the divergence is replaced by sign
change, owing to the property of the universal $O(4)$ scaling function
\cite{engels12:_scalin_funct_of_free_energ,friman11:_fluct_as_probe_of_qcd}.
At nonzero net baryon number density, the divergence at $2n$-th order
cumulants in the chiral limit appears at $n$-th order one. The
tricritical point in the chiral limit becomes the CP, where the second order
cumulants diverge
\cite{hatta03:_univer_qcd_critic_and_tricr,sasaki07:_quark_number_fluct_in_chiral,sasaki07:_suscep_polyakov}. 
The chemical freeze-out line may locate at lower
temperature than the chiral phase boundary \cite{floerchinger12:_chemic}
such that measured fluctuations
might not reflect those at the phase transition \cite{fukushima15:_hadron}. Nevertheless, the existence
of the CP is accompanied by anomalous behavior of the cumulants such as
negative third and fourth order cumulants around the CP
\cite{asakawa09:_third_momen_of_conser_charg,stephanov11:_sign_of_kurtos_near_qcd_critic_point,skokov11:_quark_number_fluct_in_polyak}
and may lead to non-monotonic behavior of  the higher order cumulants as
functions of $\sqrt{s_{NN}}$. Indeed, the measured net-proton number
cumulants in \cite{STAR_pn_2013} seems to follow this expectation,
although still inconclusive due to uncertainty. 

The measurement of the cumulants is based on event-by-event multiplicity
distribution. Once the fluctuations are regarded as those of the grand
canonical ensemble, the multiplicity distribution can be identified with
unnormalized probability distribution. 

While the cumulants are expressed by central moments of the probability
distribution, it is convenient for theoretical studies to compute them by
differentiating the thermodynamic pressure with respect to chemical
potentials.
Recently, one of the authors (K.M.) investigated the
probability distribution of the net baryon number in models with phase
transitions
\cite{morita12:_baryon_number_probab_distr_near,morita13:_net,morita14:_critic_net_baryon_number_probab}.
It turns out that sufficient information on the tail in the probability
distribution is responsible for the critical behavior of the higher
order cumulants and that the remnant of the  $O(4)$ criticality can be characterized by
narrower tail than the corresponding reference distribution. 
In the probability distribution, such information
on the phase transition is encoded in the $N$ dependence of the
canonical partition function $Z(T,V,N)$.

Since the grand partition
function is more straightforward in relativistic quantum field theories
where the number of particles are not definite, computations of the
canonical partition function are not generally easy. 
In Ref.~\cite{hasenfratz92:_canononical},
Hasenfratz and Toussaint proposed that the canonical partition
function, $Z(T,V,N)$, is calculated through the Fourier transformation of
the grand canonical partition function,$\mathcal{Z}(T,V,\mu)$, evaluated
at pure imaginary $\mu$.   
The difficulty associated with the complex fermion determinant is replaced by the
highly oscillating integral which requires extraordinary numerical
precision \cite{morita12:_baryon_number_probab_distr_near,
fukuda15:_canon_qcd, nakamura15:_canon_qcd}.

Nevertheless, the probability distribution gives further insights into
property of the system including phase transitions. 

In Ref.~\cite{nakamura13:_probin_qcd_phase_struc_by}, one of the authors
(A.N.) pointed out that one can extract the fugacity parameter $\lambda=e^{\mu/T}$
at the chemical freeze-out and construct $Z(T,V,N)$ for the net baryon number
without any assumption on the property of equilibrium $P(N)$. Furthermore, once
$Z(T,V,N)$ is known, one can obtain the grand partition function
$\mathcal{Z}(T,V,\mu)$ as a series of fugacity. This enables us
to apply Yang-Lee theory for the phase transition
\cite{yang52:_statis_theor_of_equat_of,lee52:_statis_theor_of_equat_of}(For
recent reviews, see, e.g.,  \cite{blythe03:_lee_yang,bena05:_statis}),
in which zeros of the partition function give information on the
thermodynamic property of the system. The zeros of the partition
function are distributed on a line in the complex plane of an external
parameter and its density grows up with the system volume, then finally
coalesce into the line in the thermodynamic limit. This property 
leads us, in principle, to obtain the location and order of the phase
transition from the distribution of the zeros. 
Even in the absence of the phase transition, the zeros accumulated on 
the edge of the distribution exhibit singular behavior.
This singularity, known as Yang-Lee edge singularity
\cite{kortmann71:_densit_lee_yang,fisher78:_yang_lee_edge_singul_and},
can be regarded as a CP in the complex plane and gives influence on the
thermodynamics on the real axis \cite{ejiri14:_compl_qcd}.

In both experiments and the canonical approach in LQCD, 
$Z(T,V,N)$ at large $N$ requires such high statistics that
obtained information is limited to some finite $N$,  thus one has to truncate the
fugacity polynomial there in reconstructing the grand partition function
(See \cite{nagata12:_towar_extrem_dense_matter_lattic,nagata14:_lee_yang_qcd_rober_weiss}
for recent LQCD calculations).
It is not a priori clear whether one can obtain the correct information on
the phase transition from such a truncated partition function.
The purpose of this paper is to clarify this point.
We employ a solvable model for the chiral phase transition in QCD. 
We present the Yang-Lee zeros in a chiral random matrix model, both for
the exact grand partition function and for the reconstructed one as
a truncated fugacity series with the canonical partition function being
the coefficients.  
We discuss effects of the truncation on the distribution of the Yang-Lee
zeros and compare it with the spurious zeros
of the Skellam partition function, originated from the truncation.

In the next section, we briefly summarize the general relation among the
probability distribution, partition functions, and Yang-Lee zeros.
A chiral random matrix model and its Yang-Lee zeros are presented in
Sec.~\ref{sec:ChRM}. We demonstrate differences of  the truncation
effects on the Yang-Lee zeros between the models with and without phase
transition in Sec.~\ref{sec:truncation}.
Implications for heavy ion experiments are discussed in
Sec.~\ref{sec:discussion}.
Section \ref{sec:conclusion} is devoted to concluding remarks.
Detailed expressions for partition functions in the chiral random matrix
model are given in the Appendix.

\section{General framework}
\label{sec:general}

We start from experimentally measured data of net-baryon number
multiplicity distribution $\mathcal{P}(N)$, where $N$ is the net-baryon
number. In real experiments one measures the net-proton number
$\Delta N_p = N_p - N_{\bar{p}}$. In principle one can reconstruct
$\mathcal{P}(N)$ from $\mathcal{P}(\Delta N_p)$, $\mathcal{P}(N_p)$ and
$\mathcal{P}(N_{\bar{p}})$ \cite{kitazawa12:_reveal_baryon_number_fluct_from}.
In this study we entirely assume the isospin invariance and regard
$\mathcal{P}(\Delta N_p)$ as a proxy of $\mathcal{P}(N)$.
The shape of the distribution depends on the colliding energies,
centrality etc. The net-baryon number can take any value as long as it
can be packed within the system volume. Owing to limited statistics,
however, we do not observe such states that have too large $N$  far from its mean value $M$.
Thus, we define the possible minimum and maximum of $N_{\text{min}}$ and
$N_{\text{max}}$ as
\begin{align}
 \mathcal{P}(N < N_{\text{min}}) &= 0 \nonumber\\
 \mathcal{P}(N > N_{\text{max}})& = 0.\label{eq:p-nmax}
\end{align}

In thermal equilibrium, probability distribution of the net-baryon
number in the grand canonical ensemble reads, for the fugacity factor
$\lambda=e^{\mu/T}$, 
\begin{equation}
 P(T,V,N,\mu) = \frac{Z(T,V,N)\lambda^N}{\mathcal{Z}(T,V,\mu)},\label{eq:pofN}
\end{equation}
where $\mathcal{Z}(T,V,\mu)$ is the grand partition function
\begin{equation}
\mathcal{Z}(T,V,\mu) =  \text{Tr}[e^{-(\hat{H}-\mu \hat{N})/T}]
\end{equation}
 and $Z(T,V,N)$ is the canonical partition function
\begin{equation}
 Z(T,V,N)=\text{Tr}[e^{-\hat{H}/T}\delta_{\hat{N},N}].\label{eq:canonical}
\end{equation}
Assuming the measured multiplicity distribution is the equilibrium one,
one finds $N$ dependence of $\mathcal{P}(N)$ comes from
$Z(T,V,N)\lambda^N$. Using the charge-conjugate symmetry 
$Z(T,V,-N) = Z(T,V,N)$, one can determine $\lambda$ from
$\mathcal{P}(N)$ and obtain the canonical partition function \cite{nakamura13:_probin_qcd_phase_struc_by}
\begin{equation}
 Z(T,V,N) = \mathcal{P}(N)\lambda^{-N}.
\end{equation}
Because of limited range of $N$ \eqref{eq:p-nmax}, the canonical
partition function \eqref{eq:canonical} can be obtained for
$ N \in  [\text{max}(0,N_{\text{min}}), N_{\text{max}}]$.
For the energy scan range in RHIC experiments, $N_{\text{min}} < 0$,
\textit{i.e.}, there are a few events in which more anti-protons are
observed than protons, except for $\sqrt{s_{NN}}=7.7$ GeV where
$N_{\text{min}}=1$ \cite{STAR_pn_2013}.  Note that we need only $N \geq 0$ thanks to the
charge conjugate symmetry. Thus, in most cases, one can extract the
canonical partition function for 
$-N_{\text{max}} \leq N \leq N_\text{max}$. 

This limitation in $N$ also applies to theoretical approaches such as
model studies
\cite{morita12:_baryon_number_probab_distr_near,morita13:_net,morita14:_critic_net_baryon_number_probab}
and lattice simulations
\cite{nagata12:_towar_extrem_dense_matter_lattic,nagata14:_lee_yang_qcd_rober_weiss}.
In the former, the canonical partition functions have been calculated
using a projection formula
\begin{equation}
 Z(T,V,N) = \frac{1}{2\pi i }\oint d\lambda \frac{\mathcal{Z}(T,V,\lambda)}{\lambda^{N+1}}.\label{eq:canonical_projection}
\end{equation}
where integration contour $C$ in complex $\lambda$ plane can be arbitrary, but
it is convenient to take the unit circle $\lambda = e^{i\theta}$. Then
the formula becomes 
\begin{equation}
 Z(T,V,N) = \frac{1}{2\pi}\int_{-\pi}^{\pi} d\theta \, \cos(N\theta) \mathcal{Z}(T,V,\theta)\label{eq:projection_theta}
\end{equation}
where $\theta$ is related to an imaginary  chemical potential
$-i\mu/T$. In
Refs.~\cite{morita12:_baryon_number_probab_distr_near,morita13:_net,morita14:_critic_net_baryon_number_probab}, 
thermodynamic potential $\Omega = -pV$ in Landau theory
\cite{morita12:_baryon_number_probab_distr_near} and in chiral
quark-meson model
\cite{morita13:_net,morita14:_critic_net_baryon_number_probab}
was used through $\mathcal{Z}(T,V,\theta) = e^{-\Omega(T,\theta)/T}$.
Owing to the rapid oscillation in large $N$, it turned out that the
numerical integration in double precision works up to  $P(N)\simeq
10^{-12}$.

In lattice QCD simulations, 
two approaches can provide the canonical partition functions,
i.e., (i) the fugacity expansion of the fermion determinant \cite{nagata12:_eos_qcd_wilson_taylor}
and (ii) Hasenfratz and Toussaint method \cite{hasenfratz92:_canononical}.
In the fugacity expansion, we must diagonalize a matrix whose rank is
proportional to the lattice spacial volume.  This requires large computational
resource and currently one cannot go to simulations on large lattices. 
In the method (ii), as $N$ increases, more accuracy is needed, and 
consequently $N$ cannot go to very large.

Once the canonical partition function $Z(T,V,N)$ is obtained, one can
also construct a \textit{truncated} grand canonical partition function
as a series in $\lambda$
\begin{equation}
 \mathcal{Z}^{\text{tr}}(T,V,\lambda;N_\text{max}) = \sum_{N=-N_{\text{max}}}^{N_{\text{max}}}Z(T,V,N)\lambda^{N}.\label{eq:fugacityexp}
\end{equation}
Owing to the truncation of the series at $N=-N_{\text{max}}$ and
$N_{\text{max}}$, this partition function is an approximation of the
exact partition function  which could be obtained if one can take
$N_{\text{max}}=N^*$ with $N^*$ being the number of net-baryons
fulfilling the system volume \cite{yang52:_statis_theor_of_equat_of}. 
For lattice QCD at finite temperature  with $N_s^3\times N_t$ lattice,
$N^* = 2 N_s^3$ \cite{nagata12:_eos_qcd_wilson_taylor}\footnote{In
Ref.~\cite{nagata12:_eos_qcd_wilson_taylor}, $N^*$ was derived for the quark
fugacity series as $N_q^* = 2N_c N_s^3$}. 
Thus, one needs to establish relations of physical quantities obtained
from the truncated partition function \eqref{eq:fugacityexp} with those
from the exact partition function. As seen in  the summation running from
$-N_{\text{max}}$, relativistic partition functions  contain negative
powers of $\lambda$. The suppression of high $N$ contribution to
$\mathcal{Z}$ cannot be realized by small $\lambda$. One needs to know
large $N$ behavior in $Z(T,V,N)$. 

Similar studies on higher order cumulants of the net baryon number have
been carried out in Ref.~\cite{morita13:_net}, in which sufficiently
large $N_{\text{max}}$ depending on the order of the cumulants is shown
to be necessary to obtain a correct value of the cumulants.
In this paper, we focus
on Yang-Lee zeros for the  baryon  chemical potential. 

The zeros of the partition function in complex chemical potential plane 
can be obtained by solving an equation 
\begin{equation}
 \mathcal{Z}(T,V,\mu) = 0,
\end{equation}
for complex $\mu$.
Owing to the negative powers of the fugacity, the
equation is a polynomial one in $\lambda$ with order $2N^*$.
For the truncated partition function $\mathcal{Z}^{\text{tr}}$,
one needs to solve
\begin{equation}
 \mathcal{Z}^{\text{tr}}(T,V,\mu;N_{\text{max}})=0\label{eq:z_tr_LYZ}
\end{equation}
of which the order of the polynomial is $2N_{\text{max}}$.

In the exact case, the roots have both real and imaginary part and its
distribution in the complex chemical potential plane is expected to
form a line, which crosses the real axis at the transition point in the
thermodynamic limit. The behavior of the distribution depends on the
nature of the phase transition. In
Ref.~\cite{stephanov06:_qcd_critic_point_and_compl}, the behavior of the
Yang-Lee zeros around the CP was studied by using a chiral random
matrix model. The singularity associated with the CP appears as a
branch point in complex $\mu$ plane and its property is shown to be
connected with the universality.  In lattice QCD, the phase transition
between different $Z(N_c)$ sector in the deconfined phase, Roberge-Weiss
phase transition, has been recently analyzed from a view point of
Yang-Lee zeros \cite{nagata14:_lee_yang_qcd_rober_weiss}. In this work,
we use a chiral random matrix model similar to used in
Ref.~\cite{stephanov06:_qcd_critic_point_and_compl}  but with an
extension to periodic property in imaginary chemical potential as it is
necessary to have integer net baryon number. 

The partition function is written as a polynomial in 
$\lambda = e^{\mu/T}$.
Since a complex root $\lambda_1$ is accompanied with its conjugate
$\bar{\lambda}_1$ and the charge conjugate symmetry implies
$1/\lambda_1$ and $1/\bar{\lambda}_1$ are
also roots, only the roots located in the first quadrant of the complex
$\mu$ plane are independent. In practice, it is convenient to use
Joukowski transformation $\omega=\lambda + 1/\lambda (=2\cosh\beta\mu)$ and reorganize the series in
terms of $\omega$  to reduce the number of roots to search for.
Using a property of Chebychev polynomial $T_n(\cosh x) = \cosh(nx)$,
one finds
\begin{equation}
 \begin{split}
  \lambda^N + \frac{1}{\lambda^N} &= 2\cosh(\beta \mu N)  = 2T_N(\cosh\beta\mu) \\
  &= 2 T_N(\omega/2).
  \end{split}
\end{equation}
Then Eq.~\eqref{eq:fugacityexp} reduces to a series expression
 containing only positive powers. After expanding the Chebychev
 polynomial by Eq.~\eqref{eq:chev_exp}, one finds,
\begin{equation}
\begin{split}
 \mathcal{Z}^{\text{tr}}&(T,V,\omega;N_{\text{max}})\\
 &=Z(T,V,N=0) + \sum_{n=1}^{N_{\text{max}}}
  nZ(T,V,n) \\
 &\times\sum_{k=0}^{[n/2]}(-1)^k \frac{(n-k-1)!}{k! (n-2k)!}\omega^{n-2k}.\label{zg_truncated}
\end{split}
\end{equation}
This formula could be also useful to compare a relativistic system with
nonrelativistic ones. 
The roots of $\omega$ space is easily converted into those in $\lambda$
and $\mu$ plane as
\begin{equation}
 \frac{\mu}{T} = \pm \cosh^{-1}\frac{\omega}{2}
\end{equation}
Taking both signs, one can finds all the roots in the complex $\mu$ and
$\lambda$ plane.

\section{Chiral random matrix model}
\label{sec:ChRM}

In this section, we introduce a chiral random matrix model which is an
effective model for the spontaneous chiral symmetry breaking in
QCD. Since this model is analytically solvable in the
chiral and thermodynamic limit \cite{halasz98:_phase_qcd}
and analytic expression for the partition function in  finite volume is
known \cite{stephanov06:_qcd_critic_point_and_compl}, we find this
model as the most suitable one for the present purpose.
An apparent shortcoming of the model for applying to the net baryon
number probability distribution is lack
of periodicity in imaginary chemical potential, which is a consequence
of $U(1)_B$ symmetry. Thus, we first extend the model to exhibit the
appropriate periodicity and the phase structure in the imaginary baryon
chemical potential.

In QCD, the partition function has a periodicity
$2\pi/N_c$ in the imaginary quark chemical potential $\theta_q =
\theta/3$, thus $2\pi$ in the baryon number.  LQCD simulations have
shown that there is no phase transition
in imaginary baryon chemical potential at temperatures below chiral
crossover temperature and thermodynamic quantities smoothly behave as
$\sim\cos\theta$
\cite{forcrand02:_qcd,forcrand07:_n_qcd,d'elia03:_finit_qcd}. 
This fact combined with Eq.~\eqref{eq:projection_theta} implies that the phase
transition at large baryon number density is encoded in higher Fourier
coefficients of the smoothly oscillating function.

\subsection{Partition function and thermodynamics}

We start with a partition function of the chiral random matrix model with $N_s$
sites given in \cite{halasz98:_phase_qcd}
\begin{equation}
 \mathcal{Z}_{\text{RM}} = \int\mathcal{D}X \exp\left(
				       -\frac{N_s}{\sigma^2}
				       \text{Tr}XX^{\dagger}\right)
 \text{det}^{N_f}(D+m)
\end{equation}
where $\sigma$ denotes the variance of the random matrix $X$
which has $N_s\times N_s$ dimension and $D$ is the $2N_s\times 2N_s$
matrix approximating the Dirac operator. At $T=\mu=m=0$, $\sigma$ is the
only dimensionful parameter. We use it as a unit of mass in the model
and put $\sigma=1$ in expressions below.

The Dirac operator takes the form
\begin{equation}
 D = 
  \begin{pmatrix}
   0& iX+iC \\
   iX^\dagger + i C& 0
  \end{pmatrix}.
\end{equation}
The matrix $C$ describes the effect of temperature and chemical potential.
In Ref.~\cite{halasz98:_phase_qcd}, it was chosen as
\begin{equation}
 C_k = a\pi T + \frac{b\mu}{iN_c}\label{eq:linearansatzforc1}
\end{equation}
for one half of eigenvalues and 
\begin{equation}
 C_k = -a\pi T + \frac{b\mu}{iN_c}\label{eq:linearansatzforc2}
\end{equation}
for the other half \footnote{Note that $\mu$ is the chemical potential
of the  baryon number, thus $\mu/N_c$ stands for that of the quark
number.}, with $a$ and $b$ being the dimensionless parameters.

The linear ansatz for the matrix $C$
\eqref{eq:linearansatzforc1}-\eqref{eq:linearansatzforc2}
accounts for the fact that there are the two smallest Matsubara
frequencies $\pm\pi T$. This model does not have any
thermal distribution which gives the fugacity factor $e^{\mu/T}$
thus nor the periodicity in imaginary chemical potential, since it
appears as a result of summation over the Matsubara frequencies. In order to
make the partition function periodic, we perform a following replacement
\begin{align}
 \frac{b}{N_c}\mu + i\pi a T &= \pi a T \left( i + \frac{b}{a\pi
 N_c}\frac{\mu}{T}\right)\label{eq:originalmu-T}\\
 &\rightarrow \pi a T \left( i + \frac{b}{a \pi N_c}
 2\sinh\frac{\mu}{2T} \right)\label{eq:const_replace}
\end{align}
which gives a periodicity $2\pi T$ in $\mu_I$ to the partition function.
Compared to the original linear ansatz, this replace does not change
anything at $\mu=0$ but alters the phase structure at $\mu > 2T$.

The phase structure of the model is easily evaluated by taking 
$N_s \rightarrow \infty$ limit.  Introducing an auxiliary $N_f \times N_f$
complex matrix field $\phi$ and performing the Gaussian integration with respect
to $X$, one obtains the partition function \cite{kogut04:_phases_of_quant_chrom}
\begin{equation}
 \mathcal{Z}_{\text{RM}} = \int \mathcal{D}\phi \exp[-N_s \Omega(\phi)]\label{eq:z_saddlepoint}
\end{equation}
where $\Omega(\phi)$ stands for the effective potential. Then the
partition function can be determined by the minimum of the potential,
which is evaluated at the saddle point $\phi_0$ of the
integrand:
\begin{equation}
 \left.\frac{\partial \Omega(\phi)}{\partial \phi}\right|_{\phi=\phi_0} = 0,\label{eq:saddle}
\end{equation}
and
\begin{equation}
 \lim_{N_s \rightarrow \infty}\frac{1}{N_s}\ln \mathcal{Z}_{\text{RM}} = -\min_{\phi}\Omega(\phi).
\end{equation}
The saddle point $\phi_0$ is related to the chiral condensate through
\begin{align}
 \langle \bar{\psi}\psi \rangle &= \frac{1}{N_f V_4}\frac{\partial \ln
  \mathcal{Z}_\text{RM}}{\partial m},\\
 &= \frac{1}{N_f V_4}\frac{N_s}{\sigma}2\text{ReTr}\phi_0\label{eq:psibarpsi}
\end{align}
where the four dimensional volume $V_4$ corresponds to $N_s$ such that
$N_s$ represents the typical number of the instanton (or anti-instanton)
in $V_4$. 
For real $m$, one expects $\phi_0$ is a real matrix proportional to the
unit matrix. Therefore, the saddle point can be obtained by solving
\eqref{eq:saddle} for the potential
\begin{align}
 \Omega&(\phi)/N_f \nonumber\label{eq:rm_thermopotential}\\
 = &\phi^2 -\frac{1}{2}\ln\left\{ \left[(\phi+m)^2- \tilde{T}^2
		     \left(A\sinh\frac{\mu}{2T}+i\right)^2\right]\right.\nonumber\\
 & \qquad \left.\left[(\phi+m)^2- \tilde{T}^2 \left(A\sinh\frac{\mu}{2T}-i\right)^2\right]\right\}
\end{align}
where
\begin{equation}
 \tilde{T} \equiv \pi a T
\end{equation}
and
\begin{equation}
 A \equiv \frac{2b}{a\pi N_c}.
\end{equation}

In the chiral limit $m=0$, one finds that $\phi_0=1$ at $T=\mu=0$ and a
second order phase transition occurs at $T=1/(\pi a)$  and $\mu=0$,
where $\phi_0$ continuously approaches to zero. Thus, $\phi_0$ can be
regarded as an order parameter of the chiral phase transition.

The parameters in the model, $\sigma$, $a$ and $b$, are determined as follows.
The only dimensionful parameter $\sigma$ is estimated  to be 
$\sigma \sim 100$MeV through Eq.~\eqref{eq:psibarpsi} by putting 
$\langle \bar{\psi}\psi \rangle \sim 2~\text{fm}^{-3}$ at $T=\mu=0$.
Since $T_c = 1/(\pi a)$ at $\mu=0$,  putting $T_c = 160$
MeV yields $a = 0.2$. The remaining parameter $b$ connects the model to
the density scale. With the linear ansatz for $C$
\eqref{eq:linearansatzforc1}-\eqref{eq:linearansatzforc2}, one finds the
first order phase transition at $T=0$ and $b\mu/N_c = 0.528$. We follow
the choice of Ref.~\cite{halasz98:_phase_qcd} and put $b=0.13$,
corresponding to the first order transition point at 
$\mu_c \simeq 1200$MeV, though we do not have the same phase diagram as
Ref.~\cite{halasz98:_phase_qcd} owing to the implementation of the
periodicity \eqref{eq:linearansatzforc1}-\eqref{eq:linearansatzforc2}.

 \begin{figure}[!t]
  \centering
  \includegraphics[width=\columnwidth]{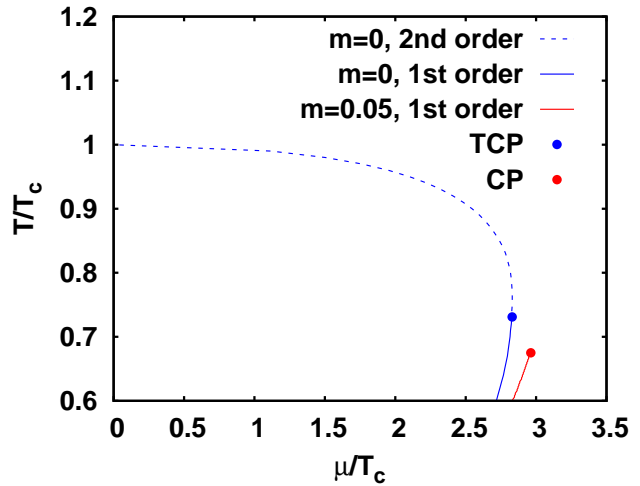}
  \caption{Phase diagram of the chiral random matrix model with
  periodicity in the imaginary baryonic chemical potential. }
  \label{fig:phasediagram}
 \end{figure}

Figure \ref{fig:phasediagram} shows the phase diagram of the modified
random matrix model \eqref{eq:rm_thermopotential} in the chiral limit
and in the presence of a small explicit symmetry breaking, $m=0.05$, respectively.
In the chiral limit, second order line continues with decreasing temperature down to  $T > T_3$ and
$\mu < \mu_3$, where $T_3 \simeq 0.731 T_c $ and $\mu_3 \simeq 4.504$ is
the location of the tricritical point (TCP). Below $T_3$, there is the
first order phase transition line. 
At finite quark mass, the second order line is replaced by smooth
crossover and TCP becomes CP with slightly decreased temperature and
increased chemical potential, $T_{\text{CP}}=0.675T_c$ and
$\mu_{\text{CP}}=4.72$, respectively. 
While these structures are the same as those in
Refs.~\cite{halasz98:_phase_qcd,stephanov06:_qcd_critic_point_and_compl},
the apparent singularity at $T=0$ in the periodic parametrization
significantly modifies the phase boundary at low temperature. 
We stress that our purpose in this paper is to explore the property of
partition function zeros rather than determination of the phase
structure.

 \begin{figure}[!t]
  \centering
  \includegraphics[width=\columnwidth]{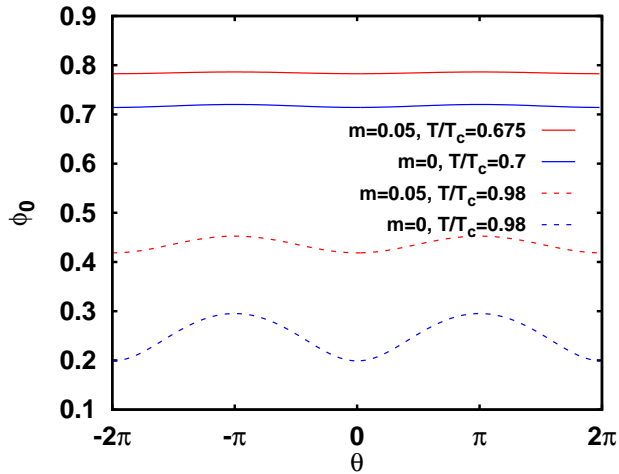}
  \caption{Behavior of order parameter $\phi_0$ of a periodic chiral
  random matrix model in imaginary chemical potential
  $\theta=\mu_I/T$ at $\mu_R=0$. }
  \label{fig:sigma-theta}
 \end{figure}

With the parameter set for $a,b$ and $\sigma$, we find that this form
also gives a reasonable thermodynamic quantities at imaginary chemical
potential.\footnote{Note that in
Refs.~\cite{halasz98:_phase_qcd,stephanov06:_qcd_critic_point_and_compl}
the coefficients in the temperature and chemical potential are absorbed
into $T$ and $\mu$. While the qualitative phase structure does not
depend on the parameters in the linear ansatz, it does so when one employs
the periodic parametrization \eqref{eq:const_replace}. }
Figure \ref{fig:sigma-theta} displays the behavior of the order
parameter $\phi_0$ in the imaginary baryonic chemical potential
$\theta=\mu_I/T$. One sees that our parameterization
\eqref{eq:const_replace} gives the correct periodicity $2\pi$ and
expected temperature dependence such as larger amplitude at higher
temperature below $T_c$
\cite{sakai08:_polyak_nambu_jona_lasin,morita11:_probin_decon_in_chiral_effec}.
Owing to lack of a $Z(3)$ sector such as the Polyakov loop background,
this model does not exhibit the Roberge-Weiss phase
transition \cite{roberge86:_gauge_qcd} at high $T$.

At finite $N_s$, the partition function can be expressed as
\cite{stephanov06:_qcd_critic_point_and_compl}
\begin{align}
 \mathcal{Z}_{\text{RM}} &=
  \sum_{k_1,k_2=0}^{N_s/2}
 \binom{N_s/2}{k_1}\binom{N_s/2}{k_2}(N-k_1-k_2)! \nonumber\\
 &\times _1\!\!F_1(k_1+k_2-N_s;1;-m^2N_s) (-N_s \tilde{T}^2)^{k_1+k_2} \nonumber \\
 &\times \left(i + A\sinh\frac{\mu}{2T}\right)^{2k_1}
  \left(i - A\sinh\frac{\mu}{2T}\right)^{2k_2}.\label{eq:grandpt}
\end{align}
where an irrelevant constant factor is ignored and $_1\!F_1(a,b;x)$
denotes the confluent hypergeometric function. One may directly obtain
zeros of this partition function, but one needs to expand $\mathcal{Z}$
in a series of the fugacity $\lambda$ to examine effects of tails in the
probability distribution function. We put the details in the Appendix
\ref{app1} and write down only the result for the canonical partition function,
 for $\delta \equiv |k_1-k_2|$,
\begin{widetext}
\begin{align}
 Z(T,N_s,N)&=\sum_{k_1,k_2=0}^{N_s/2} \binom{N_s/2}{k_1}\binom{N_s/2}{k_2}(N_s-k_1-k_2)!  _1F_1(k_1+k_2-N_s;1;-m^2N_s)  (-N_s \tilde{T}^2 A^2/4)^{k_1+k_2} \nonumber\\
 &\times \begin{cases}
	  \displaystyle \sum_{k_3=0}^{k_1+k_2} \binom{k_1+k_2}{k_3} \left[
	  \frac{2(2-A^2)}{A^2} \right]^{k_1+k_2-k_3}
	  \binom{k_3}{\frac{k_3-N}{2}}& k_1 = k_2\label{eq:zc_expanded} \\
	  \displaystyle \delta \sum_{k_3=0}^{\delta} \left(-\frac{16}{A^2}
	  \right)^{k_3} \frac{(\delta+k_3-1)!}{(\delta-k_3)! (2k_3)!}
	  \sum_{k_4=0}^{k_1+k_2-k_3} \binom{k_1+k_2-k_3}{k_4} \left[
	  \frac{2(2-A^2)}{A^2} \right]^{k_1+k_2-k_3-k_4}
	  \binom{k_4}{\frac{k_4-N}{2}}& k_1 \neq k_2  
	 \end{cases}
\end{align}
\end{widetext}

\subsection{Phase boundary and Yang-Lee zeros}

We compute the Yang-Lee zeros for the truncated partition function
\eqref{zg_truncated} with the canonical partition function of the chiral
random matrix model \eqref{eq:zc_expanded}. 
Taking $N_{\text{max}}= N_s$ in Eq.~\eqref{zg_truncated}, one recovers
the exact grand partition function \eqref{eq:grandpt}. The computation
of the zeros requires a special care in numerical digits as
cautioned in literature
\cite{halasz97:_yang_lee_qcd,nakamura13:_probin_qcd_phase_struc_by}.
We perform the calculations in 50-300 digits utilizing \texttt{FMLIB}
package \cite{FMLIB} in fortran 90.

 \begin{figure}
  \centering
  \includegraphics[width=\columnwidth]{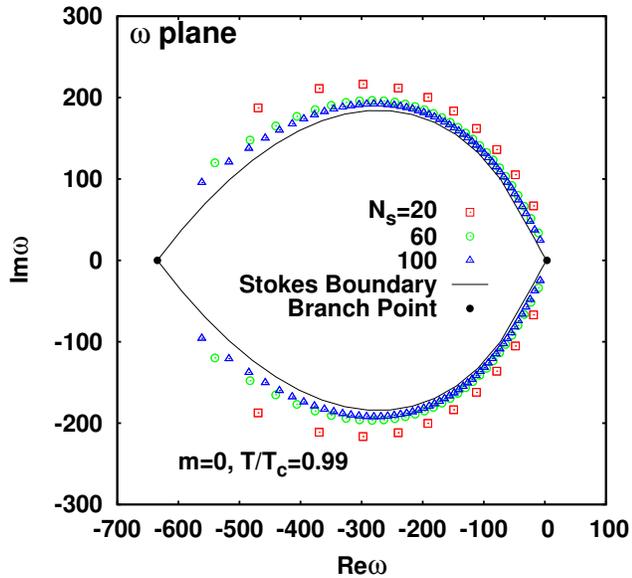}
  \caption{Yang-Lee zeros of the periodic random matrix model  in
  complex $\omega$ plane for $m=0$ at $T/T_c=0.99$. Open symbols stand
  for the zeros in different $N_s$ and  solid line indicates the Stokes
  boundary.   Branch point is denoted by  closed circles.}
  \label{fig:LYZ099_omega}
 \end{figure}

Figure \ref{fig:LYZ099_omega} shows the distribution of the Yang-Lee zero of the
periodic chiral random matrix model in the complex $\omega$ plane, for $m=0$
and at $T/T_c=0.99$. The distribution of the zeros is symmetric with respect
to the horizontal axis because the partition function is an even order polynomial 
of $\omega$ and a root has its complex conjugate. 
The solid line in Fig.~\ref{fig:LYZ099_omega} stands for Stokes boundary,
which can be regarded as an extension of the phase boundary to a complex
chemical potential plane. In the thermodynamic limit 
$N_s\rightarrow \infty$,  it satisfies
\begin{align}
 \text{Re}\left( \left.\frac{\partial^2 \Omega(\phi)}{\partial
 \phi^2}\right|_{\phi=\phi_0}\right) &> 0, \\
 \text{Re}\Omega(\phi=\phi_{0,1})&= \text{Re}\Omega(\phi=\phi_{0,2}),
\end{align}
where the first condition ensures the well-defined partition function at
the saddle point of integrand in Eq.~\eqref{eq:z_saddlepoint} and the
second condition denotes the continuity of the real part of pressure at
the boundary \cite{stephanov06:_qcd_critic_point_and_compl}. 
$\phi_{0,1}$ and $\phi_{0,2}$ stand for the two out of five solutions  of the gap
equation $\partial\Omega/\partial \phi=0$  and give the minimum of
$\text{Re}\Omega$ in both sides of the boundary, respectively.
The density of the zeros increases with $N_s$ and turns into the cut
which constitutes the Stokes boundary in the thermodynamic limit. This
is clearly seen in Fig.~\ref{fig:LYZ099_omega}. There are two branch
points located on the
real axis. Since $\omega=\lambda+1/\lambda > 0$ for real $\mu$, the one
at $\text{Re}\omega = \omega_c = 3.4 > 0$ corresponds to the second order phase transition point
in real $\mu$, while the other one, $\text{Re}\omega=-634.4$ is located on the line
$\text{Im}\mu/T = \pi$. The Stokes boundary exhibits a closed curve,
reflecting the periodicity in imaginary $\mu$ and existence of the phase
boundary at real $\mu$ axis and $\text{Im}\mu/T = \pi$.

\begin{figure*}
 \centering
 \includegraphics[width=\columnwidth]{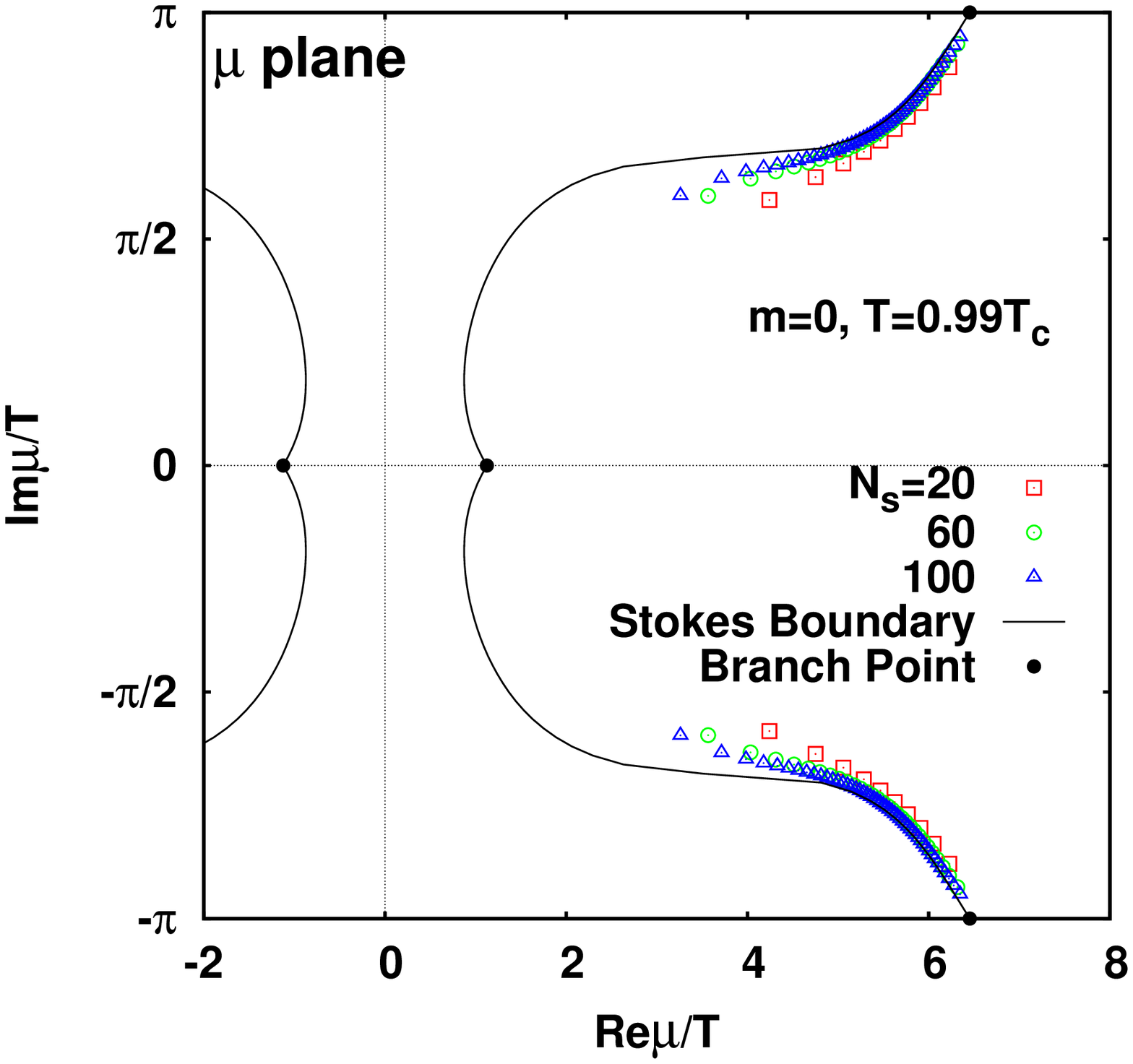}
 \includegraphics[width=\columnwidth]{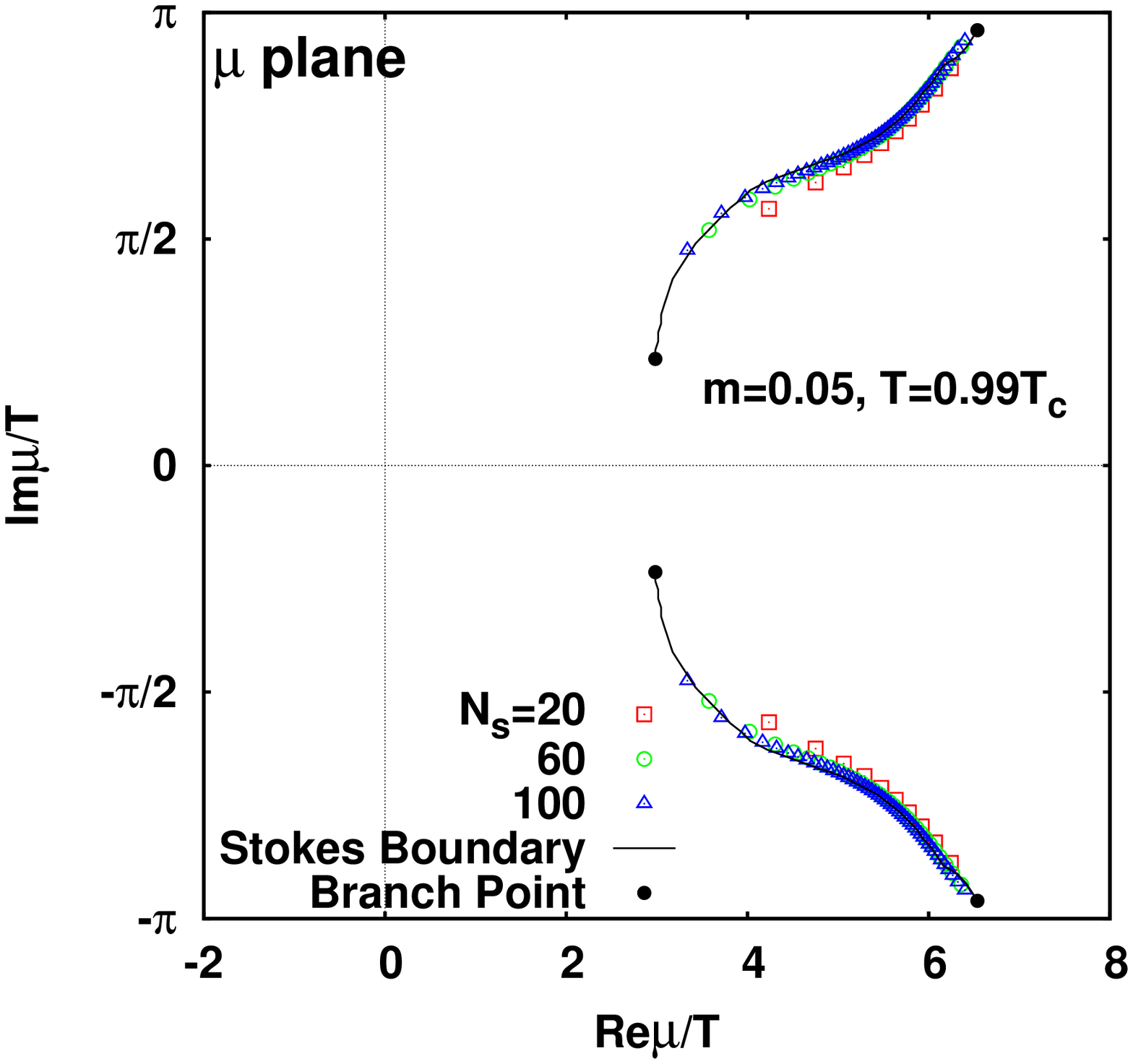}
  \caption{Yang-Lee zeros of the periodic random matrix model in complex
 $\mu$ plane. The left panel corresponds to the case of Fig.~\ref{fig:LYZ099_omega}.
  Right panel shows the case with a finite but small quark mass, $m=0.05$ at the same temperature.}
  \label{fig:LYZ099_mu}
 \end{figure*}

The phase structure can be more intuitively understood by going to
complex $\mu$ plane. 
Figure \ref{fig:LYZ099_mu}-left displays the distribution
of the same Yang-Lee zeros as in Fig.~\ref{fig:LYZ099_omega}, but the zeros
in $\text{Re}\mu < 0$ are omitted since their locations are trivial
according to the charge conjugate symmetry $\mu \rightarrow -\mu$.
The branching point on the horizontal axis indicates the second order
phase transition point. The Stokes boundary extends to both direction in
imaginary $\mu$ and ends up at the other branch point. Note that the
branch points at $\text{Im}\mu/T=-\pi$ and $\pi$ are essentially the
same because of the periodicity. We refer to
\cite{skokov11:_mappin,friman12:_phase_trans_at_finit_densit} for
behavior of the order parameter in complex $\mu$ plane and related
topics. The zeros distribute along the boundary and becomes more dense
for large $N_s$, but distance to the real axis is not so close for these
values of $N_s$. 
The behavior of the density of the zeros is related to a property of the
thermodynamic potential which can be described by an analogy to
electrostatics \cite{stephanov06:_qcd_critic_point_and_compl}.
In this case, $\text{Re}\Omega$ can be regarded as the electrostatic
potential on the $(\text{Re}\mu/T, \text{Im}\mu/T)$ plane and the normal component of
the electric field $\boldsymbol{E}=-\nabla (\text{Re}\Omega)$ to the
Stokes boundary has a discontinuity of which amount is proportional to the density of the
zeros. We confirmed that in this model these discontinuities at large
$\text{Re}\mu$, where the zeros are dense, are much larger than those at
small $\mu$, following the expectation. 
Although the density of the zeros far from the branching point is a
model-dependent feature dependent on the shape of the Stokes
boundary,  it  is governed by the universality near the branch point on
the real axis as pointed out in
Ref.~\cite{stephanov06:_qcd_critic_point_and_compl}.

Effects of the finite but small quark mass can be seen in the right panel of
Fig.~\ref{fig:LYZ099_mu} where the distribution of Yang-Lee
zeros for $m=0.05$ at the same temperature is displayed.  Owing the
explicit chiral symmetry breaking, the phase transition becomes a
crossover such that the branch point on the real axis moves to above. As
a result, there are two branch points of which are complex conjugate
each other. The same thing occurs also to the branch point at
$\text{Im}\mu/T = \pm\pi$. Here we emphasize that these complex
singularities are, albeit unphysical,  indicating existence of a chiral
phase transition in the chiral limit. These are also known as Yang-Lee edge
singularities \cite{fisher78:_yang_lee_edge_singul_and}.
The critical point at finite density (See Fig.~\ref{fig:phasediagram})
is realized by coalescence of the branch points close to real $\mu$ axis
when temperature is decreased \cite{ejiri14:_compl_qcd}. As seen in
Fig.~\ref{fig:LYZ099_mu}-right, the Yang-Lee zeros are fairly on the
boundary line and exhibit expected behaviors.

\section{Yang-Lee zeros from truncated partition functions}
\label{sec:truncation}
As described in Sec.~\ref{sec:general}, the connection of net baryon
number multiplicity distribution \eqref{eq:pofN} with the reconstructed
grand partition function \eqref{eq:fugacityexp} could potentially
enables us to extract the Yang-Lee zeros from experimental data.
Since the results presented in the previous section correspond to
$N_{\text{max}}=N^*$, i.e., no information on the exact partition
function is lost, we need to evaluate whether one can obtain the correct
distribution of the Yang-Lee zeros when the fugacity expansion is truncated.
Furthermore, even if one starts from a partition function which does not
exhibit any phase transition such as an ideal Boltzmann gas, the
truncation produces the zeros of partition function because it is a
polynomial of order $N_{\text{max}}$. In this section we investigate in
detail the effects of the truncation on the distribution of the Yang-Lee zeros.

 \begin{figure*}
  \centering
  \includegraphics[width=\columnwidth]{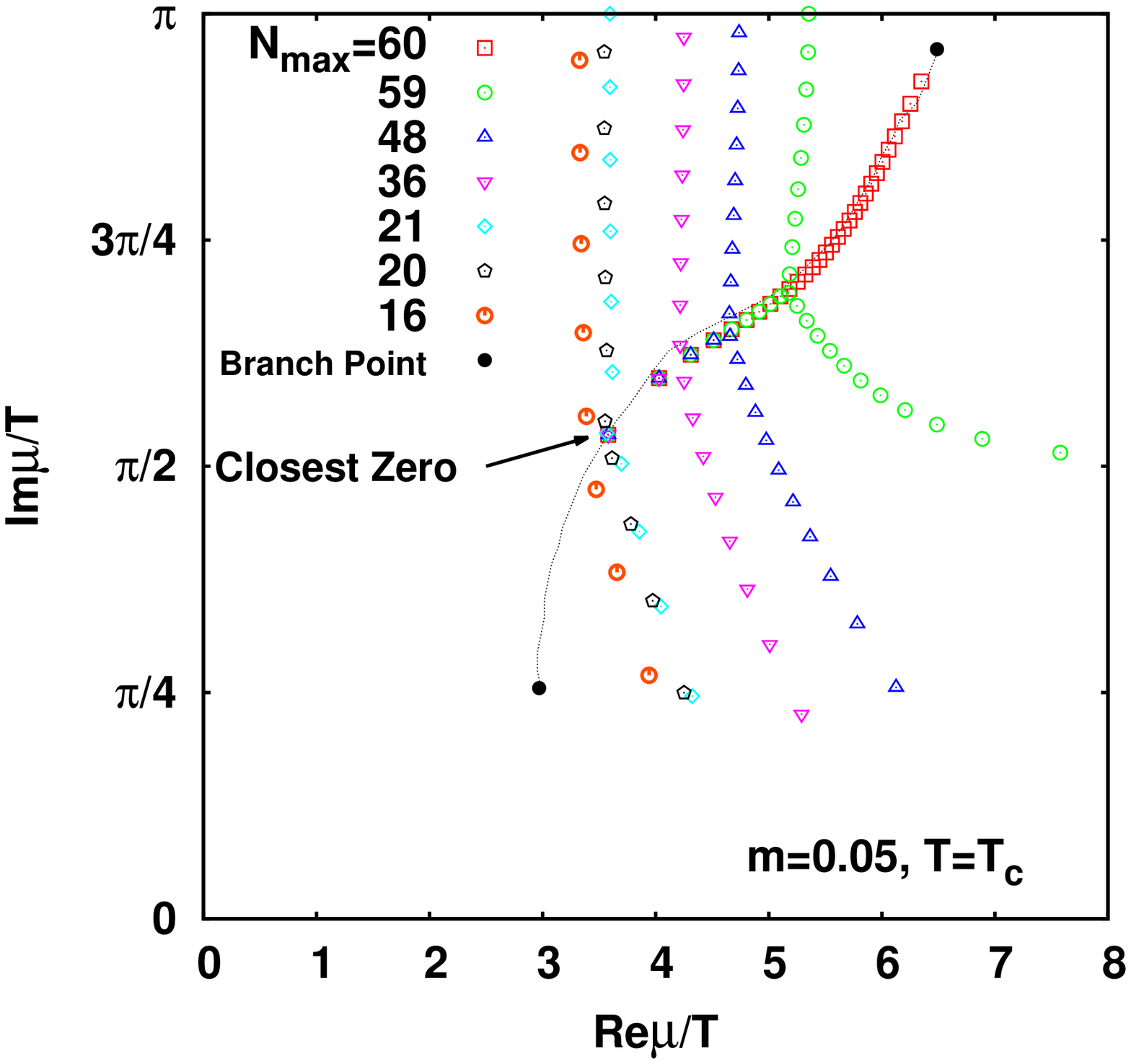}
  \includegraphics[width=\columnwidth]{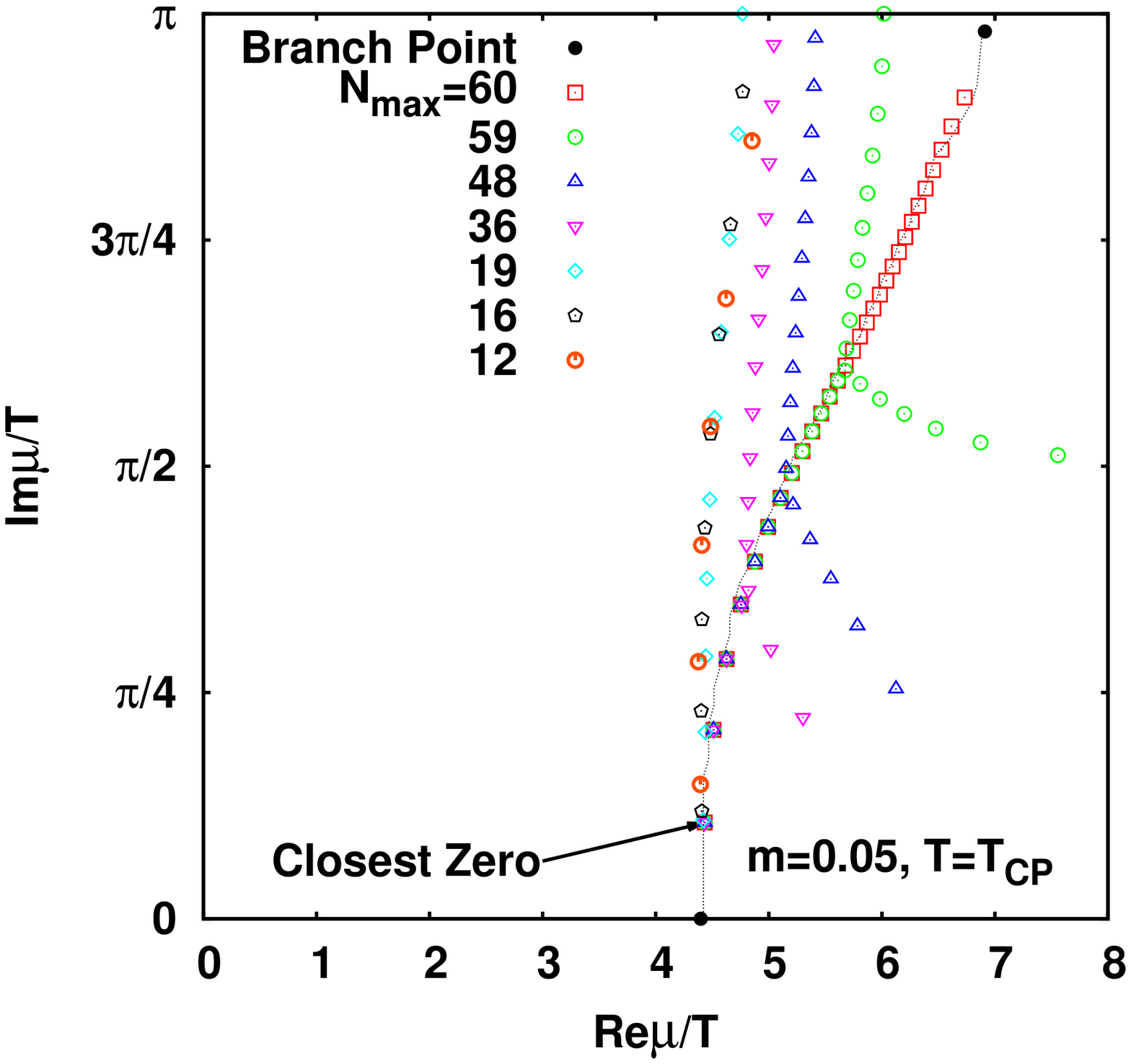}
  \caption{Distribution of the Yang-Lee zeros from the truncated partition
  function of the periodic chiral random matrix model for $m=0.05$. Left and
  right panels stand for  $T=T_c$ and $T=T_{\text{CP}}$, respectively.  }
  \label{fig:LYZ_tr1}
 \end{figure*}

\subsection{Random matrix model}

\begin{figure}
   \centering
   \includegraphics[width=\columnwidth]{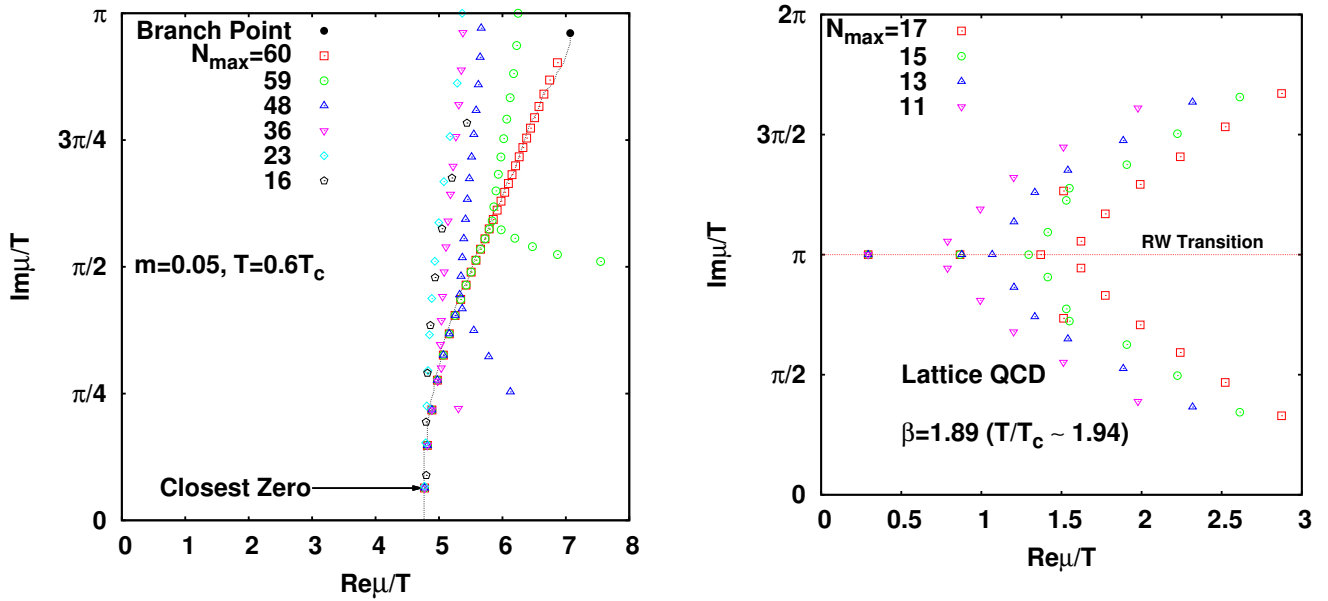}
   \caption{Same as Fig.~\ref{fig:LYZ_tr1}, but for $T=0.6T_c$}
  \label{fig:LYZ_tr_rm}
 \end{figure}

Figures \ref{fig:LYZ_tr1} and \ref{fig:LYZ_tr_rm} display the
distribution of the Yang-Lee zeros from the truncated partition function
of the periodic chiral random matrix model for various $N_{\text{max}}$
and $m=0.05$.\footnote{Note that $T_c$ is defined for $m=0$. Thus it
is slightly lower than the chiral crossover temperature for $m=0.05$.}
Hereafter we set $N_s=60$. We confirmed the following results does not
depend on the specific choice of $N_s$. We plot only the first quadrant
in complex $\mu$ plane according to the symmetry structure of the
distribution.

The left panel in Fig.~\ref{fig:LYZ_tr1} shows the case of $T=T_c$, at
which transition is of crossover type as seen in the branch point at 
$(\text{Re}\mu/T,\text{Im}\mu/T) \simeq (3, \pi/4)$. For
$N_{\text{max}}=60=N_s$, the zeros are located on the Stokes boundary
(dashed line). Reducing $N_{\text{max}}$ by one, i.e., removing
$Z(N=60)$ from the series, one sees a drastic change in the
distribution. The distribution of the zeros at large $\text{Re}\mu$ and
$\text{Im}\mu$ splits into the two lines, but the rest of the zeros remains
unchanged. Further reduction of $N_{\text{max}}$ substantially modifies
the distribution such that the splitting occurs closer to the
edge closer to the real $\mu$ axis. Nevertheless, up to $N_{\text{max}}=21$, the edge of the
distribution which is the closest Yang-Lee zero to the real $\mu$ axis remains
the same. Beyond $N_{\text{max}}=20$, the distribution no longer holds
the information on the exact Yang-Lee zeros thus the apparent relation to the phase
boundary is lost. 

The behavior with respect to changing $N_{\text{max}}$ does not depend
on temperature or corresponding phase transition. In the right panel of
Fig.~\ref{fig:LYZ_tr1}, we plot the result of the same analysis for
$T=T_{\text{CP}}=0.675T_c$ where the branch point appears on the real
axis, indicating the critical point. Reflecting the location of the
branch point, the edge of the distribution also become closer to the
real axis compared to the crossover case. The edge is stable against
decreasing $N_{\text{max}}$ down to $N_{\text{max}}=19$, then it starts
to deviate slowly when decreased further. This is so also in the case of
first order phase transition ($T=0.6T_c$) depicted in
Fig.~\ref{fig:LYZ_tr_rm}. The branch point is hidden in unphysical
Riemann sheets \cite{friman12:_phase_trans_at_finit_densit} and the edge
is very close to the real axis. 

We also note that there is always a zero at
$\text{Im}\mu/T=\pi$ when $N_{\text{max}}$ is odd. These zeros look
special since it corresponds to negative real axis in both complex
$\lambda$ and $\omega$ plane. However, this is a mathematical
consequence because in this case the truncated partition function
\eqref{zg_truncated} is an odd order polynomial, thus it has at least
one real root. As seen in  Figs.~\ref{fig:LYZ_tr1} and \ref{fig:LYZ_tr_rm}, 
it becomes the edge of one of the  lines bifurcating from the exact
Yang-Lee zeros.

These results indicate the stability of the edge does not depend on the
detail of the phase structure, although the location of the edge seems
to be connected with the shape of the Stokes boundary which is model
dependent through the $\mu$ dependence of the partition function. 
In particular, the present results are obtained by employing
the periodicity \eqref{eq:const_replace} in the random matrix model
which does not correctly take into account degrees of freedom with
baryon charges \cite{halasz98:_phase_qcd}. We note that this
modification causes unphysical behavior in thermodynamics, such as
negative $Z(T,V,N)$ at some small $N$ at low $T$, which presumably
reflect the unusual curvature of the phase boundary in Fig.~\ref{fig:phasediagram}.
Therefore, we note that the
shape of the distribution itself might not be relevant for realistic
situations. Nevertheless, below we shall see that the stability of the
edge is specific to the case with a phase transition.

\subsection{Lattice QCD}

 \begin{figure}
  \centering
  \includegraphics[width=\columnwidth]{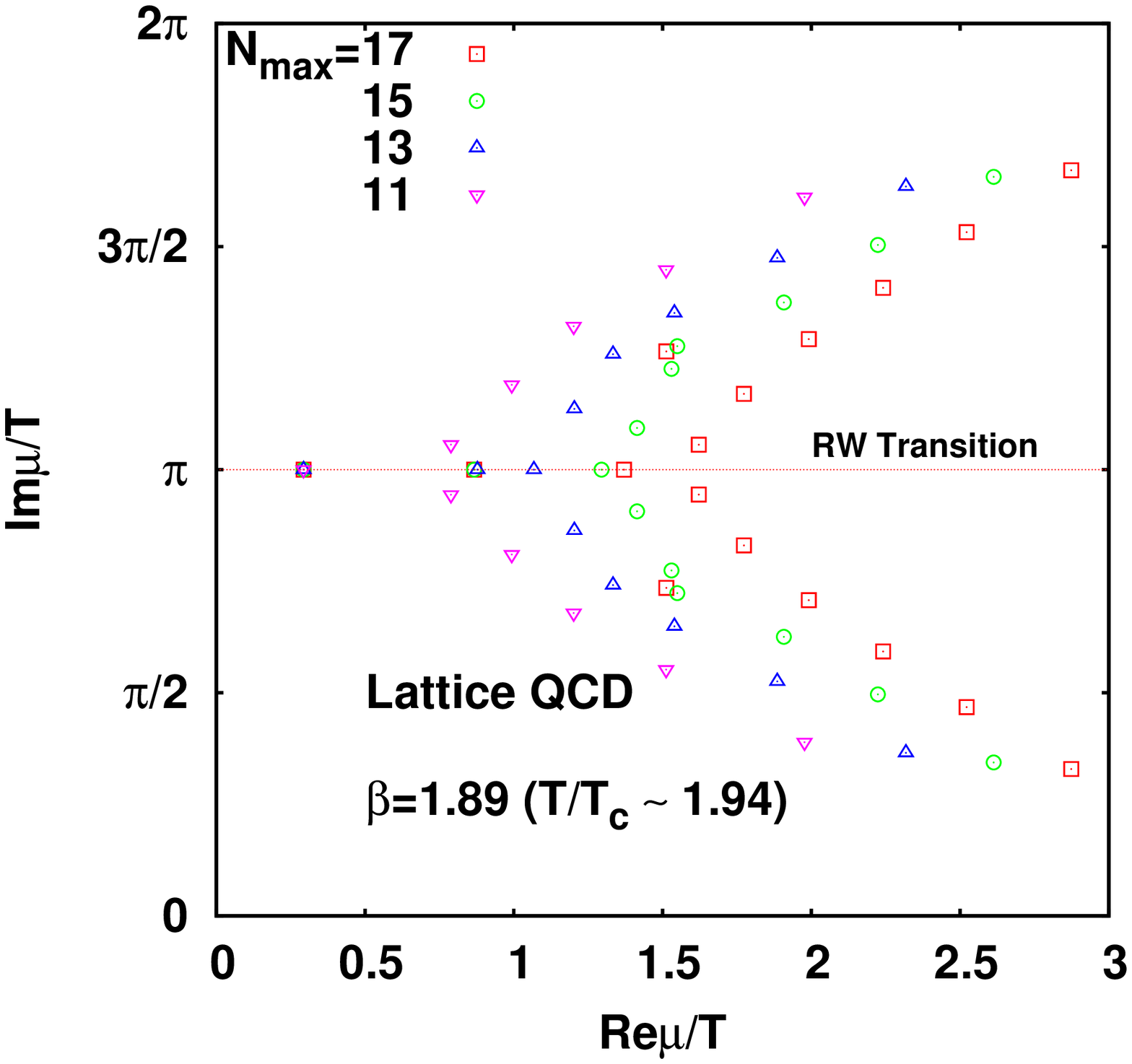}
  \caption{Distribution of Yang-Lee zeros for a lattice QCD data
  \cite{nagata12:_towar_extrem_dense_matter_lattic}.}
  \label{fig:lattice}
 \end{figure}

Figure \ref{fig:lattice} displays distribution of the Yang-Lee zeros above
$T_c$ calculated in lattice QCD simulation via the canonical method
\cite{nagata12:_towar_extrem_dense_matter_lattic}. While calculations in
the confinement phase is  still numerically difficult thus we do not see
clear indications of a phase transition at low $T$, the Roberge-Weiss (RW)
transition \cite{roberge86:_gauge_qcd} provides us a well-defined phase
transition in high temperature quark-gluon plasma phase, though at
imaginary chemical potential. In this figure, the data are calculated on
$8^3\times 4$ lattices and $\beta = 1.89$ which corresponds to
$T/T_c\simeq 1.94$.  A more detailed analysis in lattice QCD with
different lattice setups can be found in
Ref.~\cite{nagata14:_lee_yang_qcd_rober_weiss}. 
Since quark mass is heavy, the calculation is not
relevant for chiral phase transition. The RW transition is regarded as a
transition from one $Z(3)$ sector to another one when single quarks can
be excited owing to deconfinement and is known to exhibit
a first order phase transition at $\text{Im}\mu_q/T = \pm\pi/3$. In
terms of baryon chemical potential, the transition lines reduce to
$\text{Im}\mu/T = \pm\pi$ which is shown as a dotted line in
Fig.~\ref{fig:lattice}.  A brief explanation of the Roberge-Weiss phase
boundary can be found in Appendix \ref{sec:app_B}.
Since it is hard to compute $Z(N)$ near $N=N^*$,
the canonical approach in lattice QCD lacks large $N$ contribution when
one constructs the truncated partition function
\eqref{zg_truncated}. One sees that in Fig.~\ref{fig:lattice} the
behavior of the distribution of the Yang-Lee zeros against changing
$N_{\text{max}}$ is similar to that of the random matrix model, despite
the completely different origin of the phase transition. Therefore, we
expect the similar splitting behavior of the distribution also appearing
in
Refs.~\cite{nakamura13:_probin_qcd_phase_struc_by,nagata14:_lee_yang_qcd_rober_weiss}
is also due to the truncation effect. Indeed, $N_s$ and $N_{\text{max}}$
dependence of the Yang-Lee zero shown in
Ref.~\cite{nagata14:_lee_yang_qcd_rober_weiss} agrees with the
truncation effects discussed here. We expect that the bifurcation of
the zero starts at large $\text{Re}\mu$ by improving the fugacity
expansion, but one needs to take $N_{\text{max}} = N^*$ to completely
produce the Yang-Lee zero along the transition line.
In the RW transition where the transition point at $\text{Im}\mu/T =
\pm\pi$, the edge of the distribution is the closest
zero to the imaginary axis. One sees that this point is also stable
against changing $N_{\text{max}}$. This fact suggests that the stability
of the edge is not specific to the random matrix model or chiral phase
transition but might be a general property of the distribution when
partition function is truncated.

\subsection{Skellam distribution}

\begin{figure*}
 \centering
 \includegraphics[width=\columnwidth]{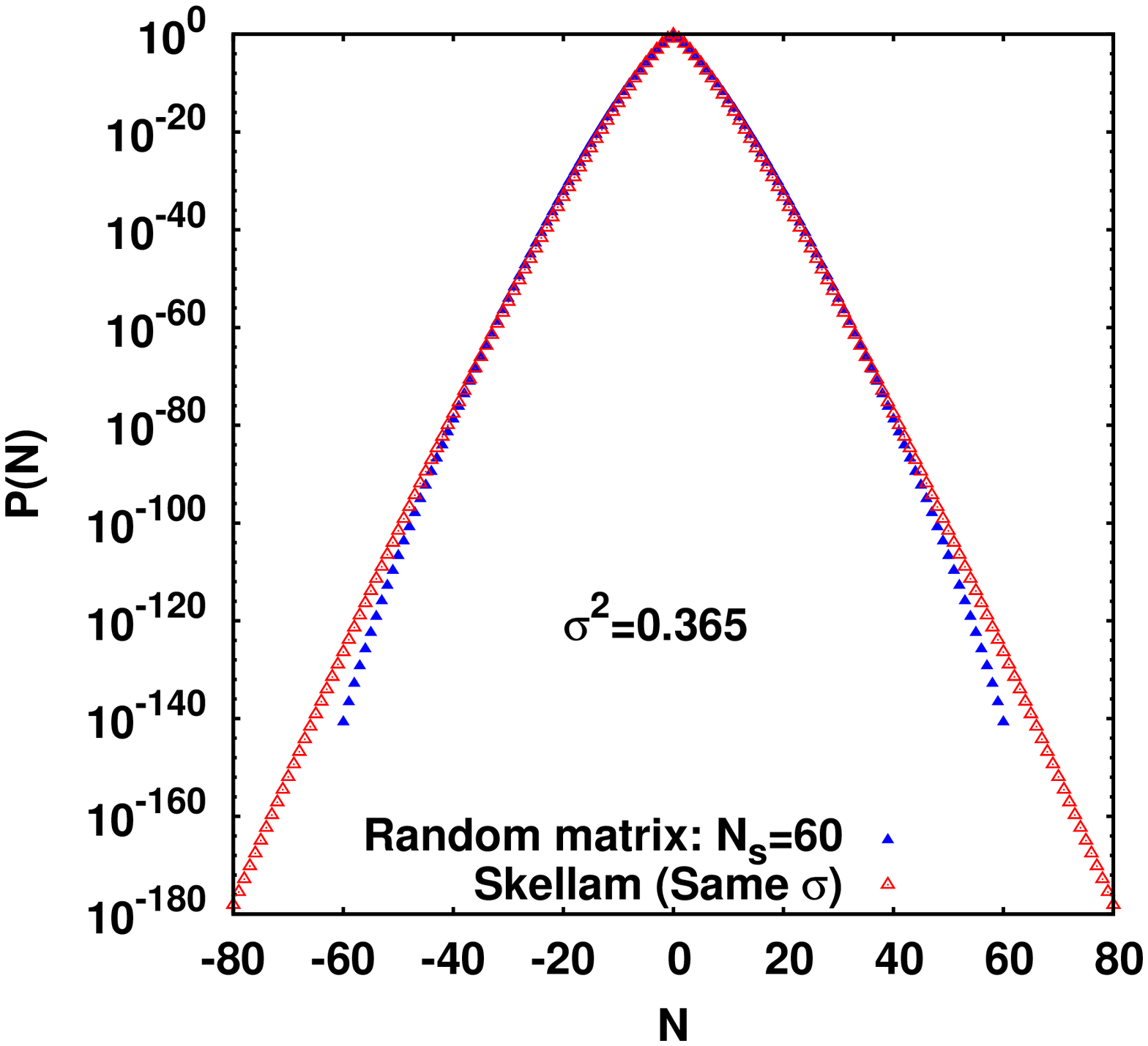}
 \includegraphics[width=\columnwidth]{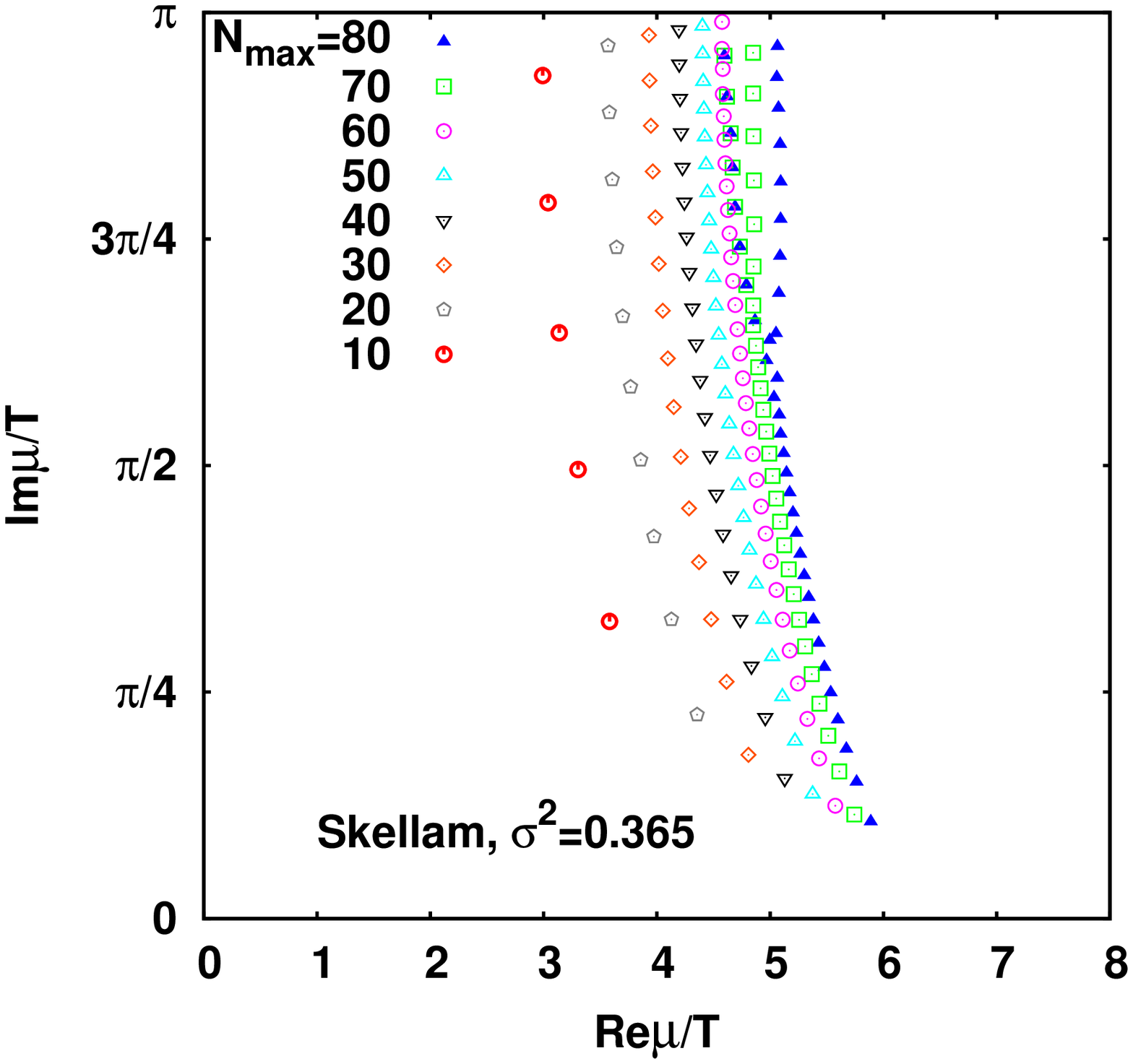}
 \caption{Left: $P(N)$ for the random matrix model at $T=T_c$ and
 $m=0.05$, and the corresponding Skellam distribution. Right:
 Distribution of Lee-Yang zeros for a truncated Skellam partition function.}
 \label{fig:skellam}
\end{figure*}

Finally we examine a model without phase transition in order to check
whether the stability of the edge is specific to phase transition or
not. We employ the Skellam distribution
\cite{skellam46:_frequen_distr_of_differ_between} of which probability
distribution of the net baryon number $N$  is given by
 \begin{equation}
  P^S(N) = \left( \frac{N_B}{N_{\bar{B}}} \right)^{N/2} I_N(2\sqrt{N_B N_{\bar{B}}}) e^{-(N_B+N_{\bar{B}})}
 \end{equation}
where $N_B$ and $N_{\bar{B}}$ denote the thermal averages of the numbers
of baryons and anti-baryons, respectively. The mean $M$ and variance
$\sigma^2$ of the distribution are given by $M=N_B-N_{\bar{B}}$ and
$\sigma^2 = N_B + N_{\bar{B}}$, respectively. For $N_B = N_{\bar{B}}$, 
The distribution becomes symmetric and the argument of the modified
Bessel function $I_N(x)$ is reduced to $2N_B = \sigma^2$.
This distribution can be derived from non-interacting Boltzmann gas
\cite{Statmodelreview_QGP3}, thus the canonical and grand canonical
partition functions read
\begin{align}
 Z(N) &= I_N(\sigma^2)\label{eq:Skellam_zc}\\
 \mathcal{Z}(\lambda)&= \exp\left[\frac{\sigma^2}{2}\left(\lambda + \frac{1}{\lambda}\right)\right]\label{eq:Skellam_zg}
\end{align}
where the temperature and volume dependence is encoded in $\sigma^2$. 
Obviously the grand partition function
\eqref{eq:Skellam_zg} does not have any roots thus no phase
transition exists. When one constructs the truncated grand partition
function \eqref{eq:fugacityexp}
from the canonical partition function \eqref{eq:Skellam_zc}, however, 
there exist complex roots. Consequently, one might see these spurious
zeros even if the system does not have any phase transition, when one
constructs the partition function through the fugacity expansion. 

Here we investigate such spurious zeros from the Skellam partition
function \eqref{eq:Skellam_zc} such that  it has the same
variance with the random matrix model at $N_s=60$, $T=T_c$ and $m=0.05$ of which
distribution of the Yang-Lee zeros is displayed in
Fig.~\ref{fig:LYZ_tr1}. Since the information on the phase transition is
encoded in the tail of the probability distribution $P(N)$, the Skellam
distribution with the same variance serves a useful reference
distribution \cite{morita14:_critic_net_baryon_number_probab}.
The probability distribution of the random matrix model and
corresponding Skellam distribution are shown in
Fig.~\ref{fig:skellam}-left. Both distributions almost agree for small
$N$, according to the same $\sigma^2$, but the deviation appears in the
tail of the distribution with tiny probability. 

Figure \ref{fig:skellam}-right shows the distribution of zeros of the
truncated partition function for the Skellam distribution with
$\sigma^2=0.365$. Except for a splitting of the
distribution for $N_{\text{max}} \geq 70$ which is similar to those in
the random matrix model, the distributions consist almost parallel lines
moving to large real $\mu$ direction as $N_\text{max}$ increases. This
behavior reflects the fact that all the zeros go away to infinity as
$N_{\text{max}}\rightarrow \infty$ since the exact grand partition
function does not have roots. Remarkably, the edges of the distributions
also move together with the rest of zeros, in contrast to the random matrix
model and lattice QCD. 
Furthermore, one notes that the distributions for
$N_{\text{max}} \leq 20$ in the random matrix model, shown in
Fig.~\ref{fig:LYZ_tr1},  resemble those from the Skellam distribution.
This observation indicates that the stability of
the edge against $N_{\text{max}}$ is a consequence of the existence of
phase transition and information on the phase transition is lost for too
small $N_{\text{max}}$.

\section{Discussion}
\label{sec:discussion}


\subsection{Comparison with $N_\text{max}$ for cumulants}
In the previous section, we have shown that the edge of the distribution
of the Yang-Lee zeros remains unchanged when the tail part of the canonical
partition function is missing. In practice, this property gives
implications for necessary statistics in heavy ion studies of the
net-baryon number fluctuations and in lattice QCD calculations. Since
the sufficient $N_{\text{max}}$
to see the stable edge depends on the system volume, here we compare it with
corresponding $N_{\text{max}}^{(i)}$ for $i$-th order cumulants. 
Here we consider only even order ones for net-baryon number at $\mu=0$,
since we are looking at $Z(T,V,N)$ rather than $P(N)$ which becomes
asymmetric with respect to $N$ at $\mu >0$. Thus the first central
moment $\delta N = N -\langle N \rangle = N$.
The second, fourth and sixth order cumulants $c_n$ $(n=2,4,6)$ read
\begin{align}
 c_2 &= \langle (\delta N)^2 \rangle \\
 c_4 &= \langle (\delta N)^4 \rangle - 3 \langle (\delta N)^2
 \rangle^2  \\
 c_6 &=  \langle (\delta N)^6 \rangle -15 \langle (\delta N)^4 \rangle
 \langle (\delta N)^2 \rangle +30 \langle (\delta N)^2 \rangle^3
\end{align}

The property of the higher
order cumulants of net-baryon number probability distribution for
changing $N_{\text{max}}$ was studied in Ref.~\cite{morita13:_net} by
using a chiral quark-meson model. For sufficiently large volume, it was
shown that $N_{\text{max}}^{(i)}$ for the cumulants approximately scale with
$\sqrt{V}$. 

 \begin{table}
  \caption{$N_{\text{max}}$ necessary for reconstructing  $i$-th order
  cumulants and edge of the Yang-Lee zeros from $Z(N)$ in the random matrix
  model at $T=T_c$ and $m=0.05$.}
  \label{tbl:nmax_vs_cumulants}
  \begin{tabular}{c|cccc}\hline
   $N_s$ & $N_{\text{max}}^{(2)}$ &
   $N_{\text{max}}^{(4)}$&$N_{\text{max}}^{(6)}$ & $N_{\text{max}}$\\ \hline
   60 & 3 & 4 & 6 & 21\\
   80 & 3& 5 & 6 & 26\\
   100& 4 & 5 & 7 & 30 \\ \hline
  \end{tabular}
 \end{table}

We summarize  the values of each $N_{\text{max}}$ in Table
\ref{tbl:nmax_vs_cumulants}. The calculations are done for $T=T_c$ and
$m=0.05$ in the random matrix model. Owing to the narrow $Z(N)$, 
even the sixth order cumulant for $N_s=100$ only requires $N_{\text{max}}^{(6)} = 7$,
i.e, $-7 \leq N \leq 7$ to reconstruct it  from $Z(T,V,N)$, while the
edge of Yang-Lee zeros demands $N_\text{max}=21$. 
The small $N_{\text{max}}^{(6)}$ implies the system volume is not large
enough to exhibit $\sqrt{V}$ scaling regime of the cumulants. This can be understood from
the small value of $\sigma^2$ in the random matrix model calculations.
The rapid decay of $P(N)$ give a rather weak dependence of
$N_{\text{max}}$ for the higher order cumulants. 
For a sufficiently large volume, one expects that $P(N)$ resembles
Gaussian near the peak, while the probability distribution
(Fig.~\ref{fig:skellam}) has a sharp peak. Thus we cannot assess the
value of $N_{\text{max}}$ needed in a realistic situation relevant for
heavy ion collisions. Moreover, the baryon number carried in this model
is not a physical one, as mentioned above. All we can say is that one
may need much more statistic than higher order cumulants. 

\subsection{Skellam distribution for large volume}
\begin{figure}
 \centering
 \includegraphics[width=\columnwidth]{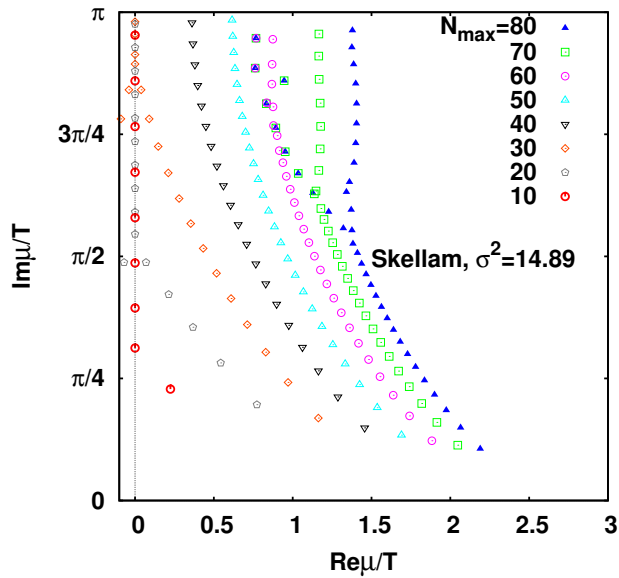}
 \caption{Distribution of the zeros for the Skellam partition function
 for $\sigma^2=14.89$. }
\end{figure}

In the fluctuation measurements at RHIC, observed $P(N)$ can be well
described by the Skellam distribution and deviation from the Skellam
distribution exists in the tail, resulting in higher order cumulants
different from the Skellam case. The obtained  variance reaches
$\sigma^2 \sim 10$
\cite{braun-munzinger11:_net_proton_probab_distr_in,STAR_pn_2013} at the
most central bin. Thus it is instructive to give a reference for the
distribution of the spurious Yang-Lee zeros based on the Skellam distribution.
Here we pick up the data for $\sqrt{s_{NN}}=7.7$ GeV at the most central
bin, which gives $\sigma^2 = 14.89$ and $M=14.41$ with available bin
from $N=0$ to $N=34$.\footnote{The data for $N=0$ and $N=34$ have only 1
event.}As mentioned in Sec.~\ref{sec:general}, one can construct $Z(N)$
from $-34 \leq N \leq 34$ according to charge conjugation symmetry
\cite{nakamura13:_probin_qcd_phase_struc_by}.
In the Skellam distribution for $\sigma^2=14.89$, we find that
$N_{\text{max}}^{(i)} = 13, 20$  and 26 for second, fourth and sixth order
cumulant, respectively. Note that these $N_{\text{max}}^{(i)}$ apply to the
cumulants at $\mu=0$. The data does not have enough statistic
for the sixth order cumulant at freeze-out $\mu$. 

We plot the distribution of the Yang-Lee zeros for the constructed Skellam
distribution with $\sigma^2=14.89$. The basic feature is the same as the
small $\sigma^2$ case (Fig.~\ref{fig:skellam}). The line of zeros moves
toward infinity as $N_{\text{max}}$ increases. One sees that some zeros
below $N=30$ appear on the imaginary $\mu$ axis, which corresponding to
the unit circle in complex fugacity plane. This means that, for a large
volume case, the zeros can appear on the imaginary axis when the tail of
the $Z(N)$ is not provided. In the Skellam distribution, these zeros can
be obtained directly by looking at the truncated partition function on
the imaginary $\mu$ axis, for $\lambda = e^{i\theta}$,
\begin{equation}
 \mathcal{Z}^{\text{tr}}_{\text{Skellam}}(\theta) =
  2\sum_{N=-N_{\text{max}}}^{N_{\text{max}}} I_N(\sigma^2) \cos N \theta,
\end{equation}
which converges into Eq.~\eqref{eq:Skellam_zg} with oscillations giving
zeros on imaginary $\mu$.

\section{Concluding remarks}

\label{sec:conclusion}
In this paper, we present analyses on partition function zeros which can
be obtained from a truncated series of the fugacity expansion. By
solving an extended chiral random matrix model which has a periodicity
in the imaginary chemical potential, we compare the exact location of
the Yang-Lee zeros and those obtained from the truncated series.
We found that the edge of the distribution of the zeros is insensitive
to the truncation of higher order terms in the fugacity expansion to
some degree. We found the similar behavior in lattice QCD at high
temperature in the context of the Roberge-Weiss phase transition.
This observation indicates that those higher order terms may have
limited influences in search for the location of the phase boundary in lattice
QCD calculations and heavy ion experiments.
 Although the distribution of
zeros exist in systems without phase transition, due to the truncation,
the zeros closest to the real $\mu$ axis are stable against truncation if
the system has a phase transition or crossover. The spurious zeros in
the Skellam distribution moves toward infinity against the truncation.
Therefore, one can distinguish whether the distribution is related to
the phase transition or not  by looking at the stability of the edge of
the distribution against the truncation. 

Although the information on the Stokes boundary is lost in the case of
too small $N_{\text{max}}$, we expect that it does not mean that all the
relevant information on the phase transition gets lost in the truncated
partition function. This expectation follows from the fact the sixth and
higher  order cumulants at $\mu=0$ should be influenced by the phase
transition and the truncated series is still able to reproduce them. 
It would be interesting to see how the distributions of the zeros in the
small $N_{\text{max}}$ cases in
Figs.~\ref{fig:LYZ_tr1}-\ref{fig:LYZ_tr_rm} are related to the remnant
of the phase transition. 

The order of the truncation in the fugacity series to obtain the stable
edge of the Yang-Lee zeros, $N_{\text{max}}$, is nevertheless found to
be much larger than those for higher order cumulants. We cannot make a
quantitative assessment on realistic values for heavy ion experiments due
to the lack of connection in the model to the real world. We hope that
such an estimate becomes feasible in the near future.

\acknowledgments
The authors would like to thank X. Luo and N. Xu for providing numerical
data of the STAR collaboration. 
They  would like to gratefully thank B. Friman and K. Redlich
for fruitful discussion and continuous encouragement.
They acknowledge stimulating discussions with Ph. de Forcrand,
F. Karsch, J.~Knoll, V.~Koch, K.~Nagata and J.~Wambach.
This work was supported by the Grants-in-Aid for Scientific Research on
Innovative Areas from MEXT (No. 24105008), by the Grants-in-Aid for
Scientific Research from JSPS (No. 15H03663, No.26610072), HIC for FAIR, and the Polish
Science Foundation (NCN), under Maestro grant 2013/10/A/ST2/00106.

\appendix

\section{Derivation of the canonical partition function in a chiral
 random matrix model}
\label{app1}

In the following, we derive an analytic expression for $Z(T,N_s,N)$ from
Eq.~\eqref{eq:grandpt}. 
First we rewrite the $\mu$ dependent part in terms of the fugacity
$\lambda = e^{\mu/T}$.

Since
\begin{align}
 \text{Re}&\left[ \left( A\sinh\frac{\mu}{2T}+i\right)^{2k_1}\left(
 A\sinh\frac{\mu}{2T}-i\right)^{2k_2} \right] \\
 &= \left( A^2 \sinh^2
 \frac{\mu}{2T} +1 \right)^{k_1+k_2}\cos[2(k_1-k_2)\phi]
\end{align}
where
\begin{equation}
 \tan\phi = \left(A\sinh\frac{\mu}{2T} \right)^{-1},
\end{equation}
and imaginary part vanishes after summation over $k_1$ and $k_2$,
using the Chebychev polynomial
\begin{equation}
 T_{k_1-k_2}(\cos2\phi) = \cos[2(k_1-k_2)\phi]\label{eq:chebychev}
\end{equation}
and
\begin{equation}
 \cos2\phi = \left( A^2 \sinh\frac{\mu}{2T} -1 \right) / \left( A^2 \sinh\frac{\mu}{2T} +1 \right),
\end{equation}
we have the partition function $\mathcal{Z}$ as
\begin{align}
 \mathcal{Z} &=
  \sum_{k_1,k_2=0}^{N_s/2} \binom{N_s/2}{k_1}\binom{N_s/2}{k_2}(N_s-k_1-k_2)!
 \nonumber \\
 &\times _1\!\! F_1(k_1+k_2-N_s;1;-m^2N_s)  (-N_s \tilde{T}^2 A^2/4)^{k_1+k_2}
 \nonumber \\
 & \times\left( \lambda + \frac{1}{\lambda} +
 \frac{2(2-A^2)}{A^2}\right)^{k_1+k_2} \nonumber \\
 &\times T_{k_1-k_2}\left( \frac{\lambda + \lambda^{-1} - 2(A^2+2)/A^2}{\lambda
 + \lambda^{-1}-2(A^2-2)/A^2} \right).\label{eq:grandpt_lambda0}
\end{align}
Expanding the Chebychev polynomial by the following expression
\begin{equation}
 T_n(x) = 
  \begin{cases}
   1 & n=0 \\
   \displaystyle n\sum_{k=0}^{n}(-2)^k
   \frac{(n+k-1)!}{(n-k)!(2k)!}(1-x)^k & n \geq 1,
  \end{cases}\label{eq:chev_exp}
\end{equation}
and using binomial expansion in the third line of \eqref{eq:grandpt_lambda0},
we can express $\mathcal{Z}$ in terms of $\lambda + \lambda^{-1}$. 
We obtain Eq.~\eqref{eq:zc_expanded} by applying the projection
\eqref{eq:canonical_projection}. 
Note that maximum power of $\lambda$  is given by $N_s$.

\section{Roberge-Weiss transition as a thermal cut}
\label{sec:app_B}

In this appendix, we give a brief explanation of the cut arising from
the Fermi distribution function and apply it to the Roberge-Weiss
transition in QCD.

\subsection{Thermal cut in free Fermi gas}

The thermodynamic potential of the free Fermi gas is given by 
\begin{equation}
 \Omega_f \sim -\int \frac{d^3 p}{(2\pi)^3} \ln[1 + e^{-\beta
  (E_p-\mu)}] + (\mu \rightarrow -\mu)
\end{equation}
where $E_p= \sqrt{p^2+m^2}$.
When the chemical potential $\mu$ has an imaginary part, 
$\mu_I = \theta T$, the imagary part gives a phase in front of
the Boltzmann factor:
\begin{equation}
 1+e^{-\beta (E_p -\mu)} = 1+e^{i\theta} e^{-\beta (E_p -\mu_R)}
\end{equation}
where $\mu = \mu_R + i \mu_I$. Therefore, for $\theta= \pm \pi$, the
phase gives $-1$ and the thermodynamic potential has a logarithmic cut
at $\theta= \pm \pi$ and $m \leq \mu_R < \infty$. The anti-particle term
also gives the cut symmetric with respect to the imaginary axis.
In Ref.~\cite{karbstein07:_how}, it is pointed out that the branch point
singularity limits the convergence radius when one tries to analytically
continue the results in the imaginary chemical potential to the real
one.
Since this cut originates from the Fermi distribution, the same analytic
structure appears in chiral models with fermions \cite{skokov11:_mappin}. 

\subsection{Roberge-Weiss transition}

In QCD at high temperature, quarks are deconfined and have a light mass
owing to chiral restoration. Since the deconfinement can be expressed as
a breaking of $Z(N_c)$ symmetry, it is useful to resort to chiral
effective models with the Polyakov loop background
\cite{fukushima04:_chiral_polyak,ratti06:_phases_qcd,schaefer10:_therm_qcd}
which successfully describe the Roberge-Weiss transition
\cite{sakai08:_polyak_nambu_jona_lasin,morita11:_probin_decon_in_chiral_effec,morita11:_role_of_meson_fluct_in}. 
Then, the relevant leading single quark contribution to the thermodynamic
potential reads
\begin{equation}
 \begin{split}
 \Omega_{q\bar{q}} \sim - \int\frac{d^3 p}{(2\pi)^3} \ln[1 + 3\Phi e^{-\beta
  (E_p-\mu_q)}]  \\ 
  + (\mu_q \rightarrow -\mu_q, \, \Phi \rightarrow  \bar{\Phi})
  \end{split}
\end{equation}
where $\mu_q = \mu/3$  is the quark chemical potential and $\Phi$ is the
the expectation value of the Polyakov loop. For antiquark contribution,
the conjugate $\bar{\Phi}$ couples to the thermal distribution. 
At the imaginary chemical potential, Polyakov loop $\Phi$ acquires a
complex phase $\varphi$.  One may express 
$\Phi = |\Phi|e^{i\varphi}$. Then the thermodynamic contribution
becomes
\begin{equation}
 1 + 3\Phi e^{-\beta(E_p-\mu_q)} = 1+3|\Phi|e^{i(\varphi - \theta_q )}
  e^{-\beta (E_q-\mu_{q,R})}.
\end{equation}

The phase of the Polyakov loop $\varphi$ varies as a function of the
imaginary quark chemical potential $\theta_q$. The Roberge-Weiss
transition at $\theta_q=\pi/3$ can be understood as a transition from
one $Z(3)$ sector with $\varphi=0$ to another one $\varphi = -2\pi/3$
\cite{morita11:_probin_decon_in_chiral_effec}. Then, the coupling
between $\varphi$ and $\theta_q$ gives the prefactor $-1$ in front of the Boltzmann
factor. Moreover, $|\Phi| \sim 1$ in the deconfined phase and the
prefactor 3 allow this function to have the singularity at 
$\mu_{q,R} = 0$.  This feature gives the cut drawn as RW transition line
in Fig.~\ref{fig:lattice}. A derivation based on the Gaussian $P(N)$ can
be found in Ref.~\cite{nagata14:_lee_yang_qcd_rober_weiss}.
In the confinement phase where $|\Phi|\sim 0$, this term is suppressed and the
thermal cut from the quark does not appear. 

%

\end{document}